\documentclass{aa}  

\def\edit
\usepackage{graphicx}
\usepackage{txfonts}

\usepackage{amsmath}
\usepackage{amssymb}
\usepackage{tabularx}
\usepackage{txfonts}
\usepackage{longtable}
\usepackage{textcomp}
\usepackage{hhline}
\usepackage{arydshln}
\usepackage{multirow}
\usepackage{lscape}
\usepackage{array}
\usepackage{natbib}
\usepackage{setspace}
\usepackage{multicol}
\usepackage{pifont} 
\usepackage{wrapfig}
\usepackage{amsmath}
\usepackage{hyperref}
\usepackage{fancybox}
\usepackage{color}
\usepackage{subcaption}
\usepackage[dvipsnames]{xcolor}
\usepackage[Export]{adjustbox}
\usepackage[normalem]{ulem}
\usepackage{siunitx}
\usepackage{acronym}
\usepackage[capitalise]{cleveref}

\usepackage{titlesec}

\Crefname{section}{Sect.}{Sects.}

\newacro{nir}[NIR]{near-infrared}
\newacro{mir}[MIR]{mid-infrared}
\newacro{fir}[FIR]{far-infrared}
\newacro{ir}[IR]{infrared}
\newacro{mm}[mm]{millimetre}
\newacro{sub-mm}[sub-mm]{sub-millimetre}
\newacro{cm}[CM]{core-mantle}
\newacro{emt}[EMT]{effective medium theory}
\newacro{dda}[DDA]{discrete dipole approximation}
\newacro{ism}[ISM]{interstellar medium}
\newacro{dlp}[DLP]{degree of linear polarisation}

\def\eff{\mathrm{eff}}
\def\min{\mathrm{min}}
\def\max{\mathrm{max}}
\def\abs{\mathrm{abs}}
\def\sca{\mathrm{sca}}
\def\ext{\mathrm{ext}}
\def\pol{\mathrm{pol}}
\def\d{\mathrm{d}}

\def\sph{Sph. }
\def\prol{Prol. }
\def\por{P.Sph.$_{0.54}$ }
\def\agg{Agg.$^{\qty{100}{\nm}}_{\qty{1}{\um}}$ }
\def\aggdemi{Agg.$^{\qty{50}{\nm}}_{\qty{1}{\um}}$ }
\def\aggice{Agg.:ice$^{\qty{50}{\nm}}_{\qty{1.26}{\um}}$ }

\makeatletter
\newlength{\sfp@hseplen}\newlength{\sfp@vseplen}
\define@cmdkey{subfigpos}[sfp@]{pos}[ul]{}% \sfp@pos
\define@cmdkey{subfigpos}[sfp@]{font}[\small]{}% \sfp@font
\define@cmdkey{subfigpos}[sfp@]{vsep}[2\baselineskip]{\setlength{\sfp@vseplen}{\sfp@vsep}}% \sfp@vsep
\define@cmdkey{subfigpos}[sfp@]{hsep}[10pt]{\setlength{\sfp@hseplen}{\sfp@hsep}}% \sfp@hsep
\newcommand{\subfigimg}[4][,]{%
  \setkeys{subfigpos}{pos,font,vsep,hsep,#2}% seulement les clés locales (font, pos, vsep, hsep)
  \setbox1=\hbox{\adjincludegraphics[#1]{#4}}% passe les options ici
  \ifnum\pdfstrcmp{\sfp@pos}{ul}=0% UPPER LEFT placement of subfig label
    \leavevmode\rlap{\usebox1}% Print image
    \rlap{\hspace*{\sfp@hsep}\raisebox{\dimexpr\ht1-\sfp@vsep}{\sfp@font{#3}}}% Print label
    \phantom{\usebox1}% Insert appropriate spacing
  \else\ifnum\pdfstrcmp{\sfp@pos}{ur}=0% UPPER RIGHT placement of subfig label
    \leavevmode\usebox1% Print image
    \llap{\raisebox{\dimexpr\ht1-\sfp@vsep}{\sfp@font{#3}}\hspace*{\sfp@hsep}}% Print label
  \else\ifnum\pdfstrcmp{\sfp@pos}{lr}=0% LOWER RIGHT placement of subfig label
    \leavevmode\usebox1% Print image
    \llap{\raisebox{\sfp@vsep}{\sfp@font{#3}}\hspace*{\sfp@hsep}}% Print label
  \else% Assume LOWER LEFT placement of subfig label
    \leavevmode\rlap{\usebox1}% Print image
    \rlap{\hspace*{\sfp@hseplen}\raisebox{\sfp@vsep}{\sfp@font{#3}}}% Print label
    \phantom{\usebox1}% Insert appropriate spacing
  \fi\fi\fi
}
\makeatother

\begin{document}

   \title{From small dust to micron-sized aggregates: the influence of structure and composition on the dust optical properties}
   \subtitle{}

   \titlerunning{From small dust to $\mathrm{\mu m}$-sized aggregates: influence of structure and composition on dust optical properties}

   \author{M.-A. Carpine\inst{1} \and
           N. Ysard\inst{2,3} \and A. Maury\inst{4,5,1}  \and A. Jones\inst{3}} %
           
   \institute{ Universit\'{e} Paris-Saclay, Universit\'{e} Paris Cit\'{e}, CEA, CNRS, AIM, 91191, Gif-sur-Yvette, France
         \\
         \email{marie-anne.carpine@cea.fr}
         \and IRAP, CNRS, Universit\'{e} de Toulouse, 9 avenue du Colonel Roche, 31028 Toulouse Cedex 4, France
         \and
         Universit{\'e} Paris-Saclay, CNRS, Institut d'Astrophysique Spatiale, 91405, Orsay, France
         \and
         Institute of Space Sciences (ICE), CSIC, Campus UAB, Carrer de Can Magrans s/n, E-08193 Barcelona, Spain
         \and ICREA, Pg. Llu\'{i}s Companys 23, Barcelona, Spain
             }

   \date{\today}

  \abstract
   {Models of astrophysical dust are key to understand several physical processes, from the role of dust grains as cooling agents in the ISM to their evolution in dense circumstellar disks, explaining the occurrence of planetary systems around many stars. Currently, most models aim at providing optical properties for dust grains in the diffuse ISM, and many do not account properly for complexity in terms of composition and structure when dust is expected to evolve in dense astrophysical environments. }
   {Our purpose is to investigate, with a pilot sample of micron-size dust grains, the influence of hypothesis made on the dust structure, porosity, and composition when computing the optical properties of grown dust grains. We aim to produce a groundwork for building comprehensive yet realistic optical properties which accurately represent dust grains as they are expected to evolve in the dense clouds, cores, and disks. We are especially interested in exploring these effects on the resulting optical properties in the infrared and millimetre domains, where observations of these objects are widely used to constrain the dust properties.}
   {Starting from the small dust grains developed in the THEMIS\,2.0 model, we use the Discrete Dipole Approximation to compute the optical properties of \qty{1}{\um} grains, varying the hypotheses made on their composition and structure. We look at the dust scattering, emission and extinction to isolate potential simplifications and unavoidable differences between grain structures.}
   {We note significant differences in the optical properties depending on the dust structure and composition. Both the dust structure and porosity influence the dust properties in infrared and millimetre ranges, demonstrating that dust aggregates cannot be correctly approximated by compact or porous spheres. In particular, we show that the dust emissivity index in the millimetre can vary with fixed grain size.}
   {Our work sheds light on the importance of taking the dust structure and porosity into account when interpreting observations in astrophysical environments where dust grains may have evolved significantly. For example, measuring the dust sizes using the emissivity index from millimetre observations of the dust thermal emission is a good but degenerate tool, as we observe up to $25\%$ differences in the dust emissivity index with compact or aggregate grains, varying composition and structure. Efforts in carrying out physical models of grain growth, for instance, are required to establish realistic constraints on the structure of grown dust grains, and will be used in the future to build realistic dust models for the dense ISM.}


   \maketitle
%
%-------------------------------------------------------------------
\section{Introduction}
\label{sec:intro}

Interstellar dust is of great influence in many astrophysical processes. 
The correct understanding and modelling of dust is undeniably necessary, whether it is to understand the formation of planetesimals \citep{Birnstiel2016}, the formation of complex molecules on their surfaces \citep{herbst_unusual_2021,dulieu_experimental_2010,Wakelam2017}, or to correctly interpret astronomical observations using dust as a tracer, in most spectral range \citep{galliano_interstellar_2018}.
In terms of physical processes, dust plays a crucial role in the \ac{ism} evolution, especially in star forming regions where it interacts with the molecular gas, but also becomes key to setting the coupling with the magnetic field in places where the ionisation fraction of the gas by cosmic rays is very low \citep{zhao_formation_2020, maury_recent_2022}, and is the core of heating and cooling processes. More importantly, dust grains constitute the building blocks of planet formation \citep[see e.g.][and references therein]{testi_dust_2014}. From an observational point of view, dust represents a major tracer, and unavoidable asset to probe the cold Universe. Dust optical properties impact light absorption, scattering, emission and polarisation, and are used to estimate various physical properties such as the gas temperature, mass and spatial distribution. Conversely, little is known about the microscopic properties of the astrophysical dust and inferring them from observations is also not straightforward.
From the diffuse \ac{ism} where only the smallest submicronic grains are observed \citep{mathis_size_1977, hirashita_synthesized_2013} to the circumstellar disks where km-sized planetary bodies are formed \citep{keppler_discovery_2018}, dust grains are expected to undergo an evolution, in terms of size and structure, as they are incorporated into astrophysical environments of increasing gas density: if small \ac{ism} grains can be considered as compact spheres or spheroids, dust growth by ballistic aggregation implies that larger grains may naturally inherit an aggregated structure. 
While the evolution of small dust into large pebbles in disks has long been explored because of its direct impact on the formation of rocky planets \citep{birnstiel_dust_2024,testi_dust_2014,teiser_growth_2025,drazkowska_how_2021}, the early evolution of dust grains in the dense ($n \sim \qty{e6}{\per\cm\cubed}$) star-forming material that makes up the pristine disk material, is still largely unexplored \citep{priestley_efficiency_2021, hirashita_condition_2013}. However, recent studies suggest that micron-sized grains may be present in dense clouds \citep{pagani_ubiquity_2010, dartois_spectroscopic_2024}, and that the material surrounding young protostars (protostellar envelopes) could host dust grains a few tens of microns in size \citep{galametz_low_2019,cacciapuoti_faust_2023}. Moreover, fractal aggregates and irregular solid grains have been observed in cometary dust in the Solar System, and are consistent with models of early dust aggregation from pristine material, in the solar nebula \citep[e.g.][]{fulle_fractal_2017}. Efforts to model complex dust particles should therefore be made not only to describe dust evolution towards planetary systems in disks, but also to interpret dust observations in dense clouds, cores, and protostellar envelopes. It is thus key to build realistic models of micron-sized grains to interpret the dust observations in such environments, and to ensure that these dust models are physically motivated by (i) constraints on dust properties in the diffuse ISM and (ii) predictions from dust growth models.

A dust model is first defined from the dust optical properties, which translate into the absorption, scattering and extinction efficiencies of the dust grains \citep[respectively $Q_{abs}$, $Q_{sca}$ and $Q_{ext}$, extensively defined in][]{bohren_absorption_1983}, tabulated by wavelength and grain size, and which can be used as an input for radiative transfer simulations \citep{reissl_radiative_2016, dullemond_radmc-3d_2012}. In observations, dust emission is driven by thermal effects, modulated by the dust emissivity, i.e. the absorption efficiency, whereas light extinction is driven by the extinction efficiency, which combines the effects of both absorption and scattering events. 
Many dust grain models have been defined over the years, using different optical properties for the silicate and carbonaceous dust components \citep[e.g.][]{ossenkopf_dust_1994, weingartner_dust_2001, compiegne_global_2011, jones_evolution_2013, guillet_dust_2018,
Siebenmorgen2023,
hensley_astrodustpah_2023, ysard_themis_2024}. 
We choose to use laboratory measurements of \ac{ism} solid matter analogues made by \citet{demyk_low-temperature_2017} which permitted to derive optical constants for amorphous silicates \citep{demyk_low-temperature_2022} and those of the hydrogenated amorphous carbons defined by \citet{jones_variations_2012, jones_variations_2012-1,jones_variations_2012-2}. The THEMIS 2.0 model \citep{ysard_themis_2024} is based on these optical constants, and fitted to observational constraints on emission and extinction in both total and polarised light in the diffuse \ac{ism}. If this model is unique in reproducing dust emission and extinction from recent observations of the diffuse \ac{ism}, it remains to be done to extend this model of dust grains to particles more representative of dust as it evolves in the densest parts of the ISM.

The first THEMIS\footnote{A full description of the model and the associated bibliography are available here: \url{https://www.ias.u-psud.fr/themis/}.} dust model \citep{jones_evolution_2013, jones_global_2017, kohler_dust_2015} has been studied in detail in terms of structure \citep{ysard_optical_2018} and composition \citep{ysard_grains_2019}. However the THEMIS 2.0 model has only been developed for the small grains typical of the diffuse ISM, and has proven successful in reproducing several features observed in this medium \citep{ysard_themis_2024}. 
We present here a first pilot sample of micron-sized grains, built from the THEMIS 2.0 small dust grains, combining for the first time complex composition and structure for this new grain model.

Using the materials defined in THEMIS 2.0, we develop a first pilot sample of micron-sized grains whose shapes vary from the simple sphere to the most complex aggregates, and study how the hypothesis made to simplify the structure of aggregates can affect dust emission, extinction, but also light polarisation.\\
This paper is organised as follows : in \cref{sec:model} we define in detail the dust grains studied and the methods used to derive the dust optical properties. In \cref{sec:results} we present the results and the variation of the different optical properties of the individual grains. We discuss and interpret these results in \cref{sec:discussion}. Lastly, we summarise the main results in \cref{sec:concl}.

%--------------------------------------------------------------------
\section{Models of micron-sized dust}
\label{sec:model}

Whereas the new silicate optical properties included in the THEMIS 2.0 dust models proposed in \citet{ysard_themis_2024} seem to capture well the dust properties observed in the diffuse medium, these types of grains have not yet been tested for reproducing dust observations in the dense medium. This work is a first effort to model the optical properties of THEMIS 2.0 grains as they evolve and grow from the diffuse ISM to the higher gas densities where stars and disks will form. We explore the properties of a first sample of micron-sized grains, focusing on two wavelength domains where observations can be made: the \ac{ir} and \ac{mm} ranges.

We have chosen to explore a limited parameter space of grain shapes and structures to carry out the first pilot models of micrometric dust grains based on THEMIS 2.0. Our sample includes: a compact spherical grain, a compact prolate grain, a porous spherical grain, and three aggregates of equivalent size $\sim 1\mu$m (see \cref{table:grains}). The calculation of the optical properties of complex and larger dust aggregates requires very large computational times: while this study consists in the first attempt to provide optical properties for physically motivated micron-sized grains made of THEMIS 2.0 material, the properties of more complex grains, expected to form in the circumstellar environments, will be explored in a forthcoming study. In the following, we develop the specific properties in terms of the composition and shape of the grains.

\subsection{Grain composition}
\label{subsec:opt.const}

Astrophysical dust is classically made out of two main components : carbonaceous material (in the form of graphite or amorphous carbon), and silicate material of varying stoichiometry.

Starting with the THEMIS 2.0 model, we use as a basis the "silicate grains" a-Sil. Amorphous a-Sil grains contain metallic iron and iron sulfide nano-inclusions, and are coated with a 5 nm thick aromatic-rich a-C carbon mantle.

The composition of the THEMIS 2.0 amorphous carbon materials of the a-C mantle are defined in \citet{jones_variations_2012,jones_variations_2012-1,jones_variations_2012-2} and remain unchanged since then. 
Regarding the silicate core; it is a combination of three different amorphous silicate samples, which were measured in the laboratory \citep{demyk_low-temperature_2017, demyk_low-temperature_2022}: two samples with an enstatite stoichiometry, and one with a forsterite stoichiometry. \citet{ysard_themis_2024} proposed different combinations or "mixes" of these three materials, and showed that they were all compatible with \ac{ism} observations. Note that, depending on the mixture, the millimetre spectral index of the silicate grains ranges from $\beta_{\mathrm{mm}} \sim1.7$ to $\sim2.25$. In this study, we examine two specific mixes, varying according to their forsterite-to-enstatite ratio: a-Sil2 and a-Sil7. These mixes correspond to the extreme values of $\beta\sim2.25$ and $\beta\sim1.7$, respectively.

For the sake of simplification, we use the \ac{emt} approximations from \citet{ysard_themis_2024} for the refractive index of a-Sil grains. This is valid for the size of monomers we are considering, whose mass fraction is largely dominated by silicates. Therefore, a single refractive index is used for the entire grain (a-Sil core and a-C mantle), and we can omit the constraints from the \ac{cm} structure. This relaxes the constraints on dipole sizes of the \acl{dda} modelling (see \cref{subsec:dda}).

The \aggice grain, described in \cref{subsec:grainshape}, is coated with a layer of pure water ice. Following the work of \citet{kohler_dust_2015}, we use the optical constants from \citet{warren_optical_1984} for water ice.

\subsection{Grain shape and structure}
\label{subsec:grainshape}

\begin{figure*}[h]
\centering
\begin{subfigure}{.33\textwidth}
  \centering
  \adjincludegraphics[width=\linewidth,trim={{.08\width} {.13\width} {.08\width} {.13\width}},clip]{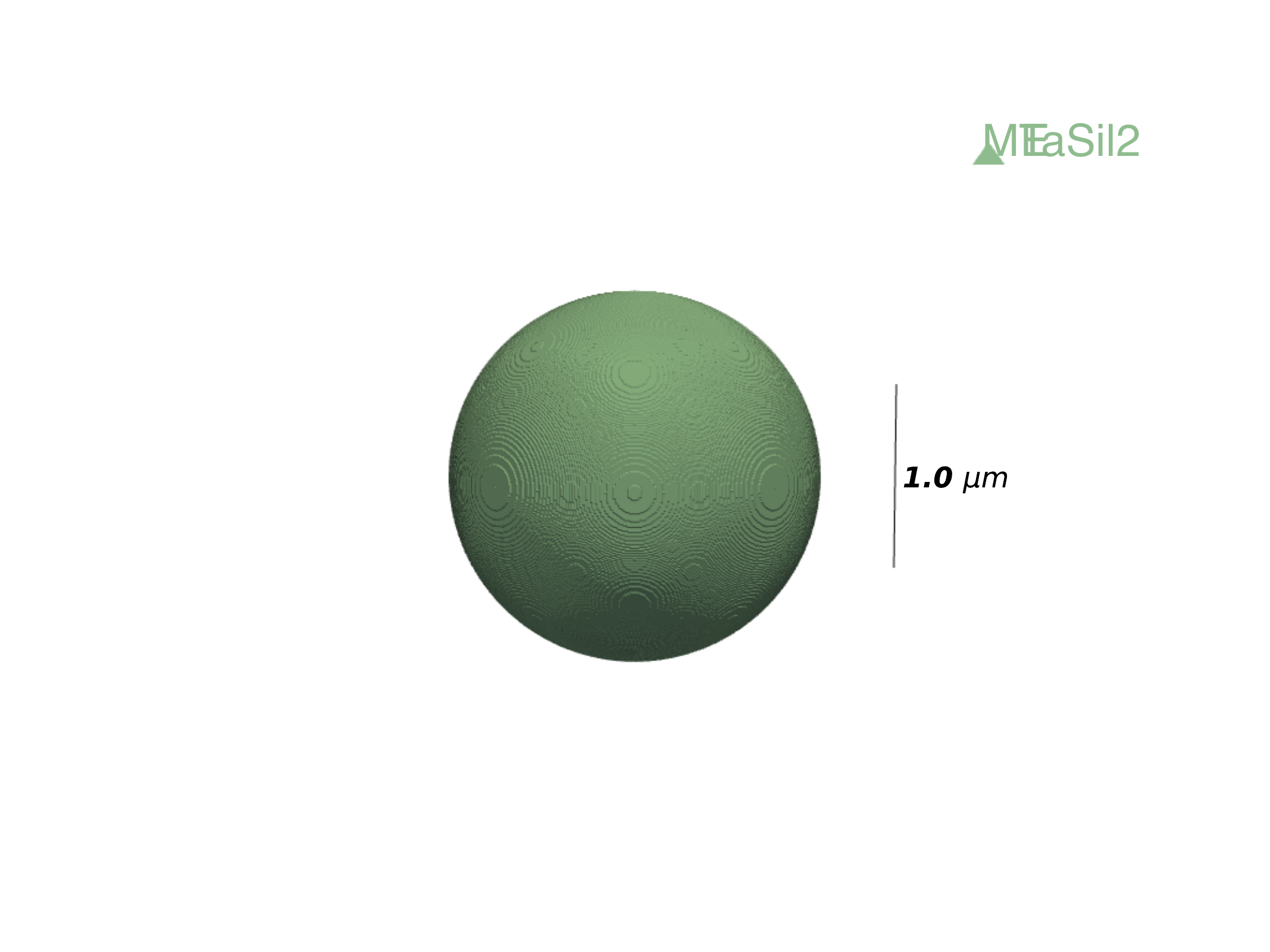}
  \caption{Sph.}
  \label{fig:sfig11}
\end{subfigure}%
\begin{subfigure}{.33\textwidth}
  \centering
  \adjincludegraphics[width=\linewidth,trim={{.08\width} {.13\width} {.08\width} {.13\width}},clip]{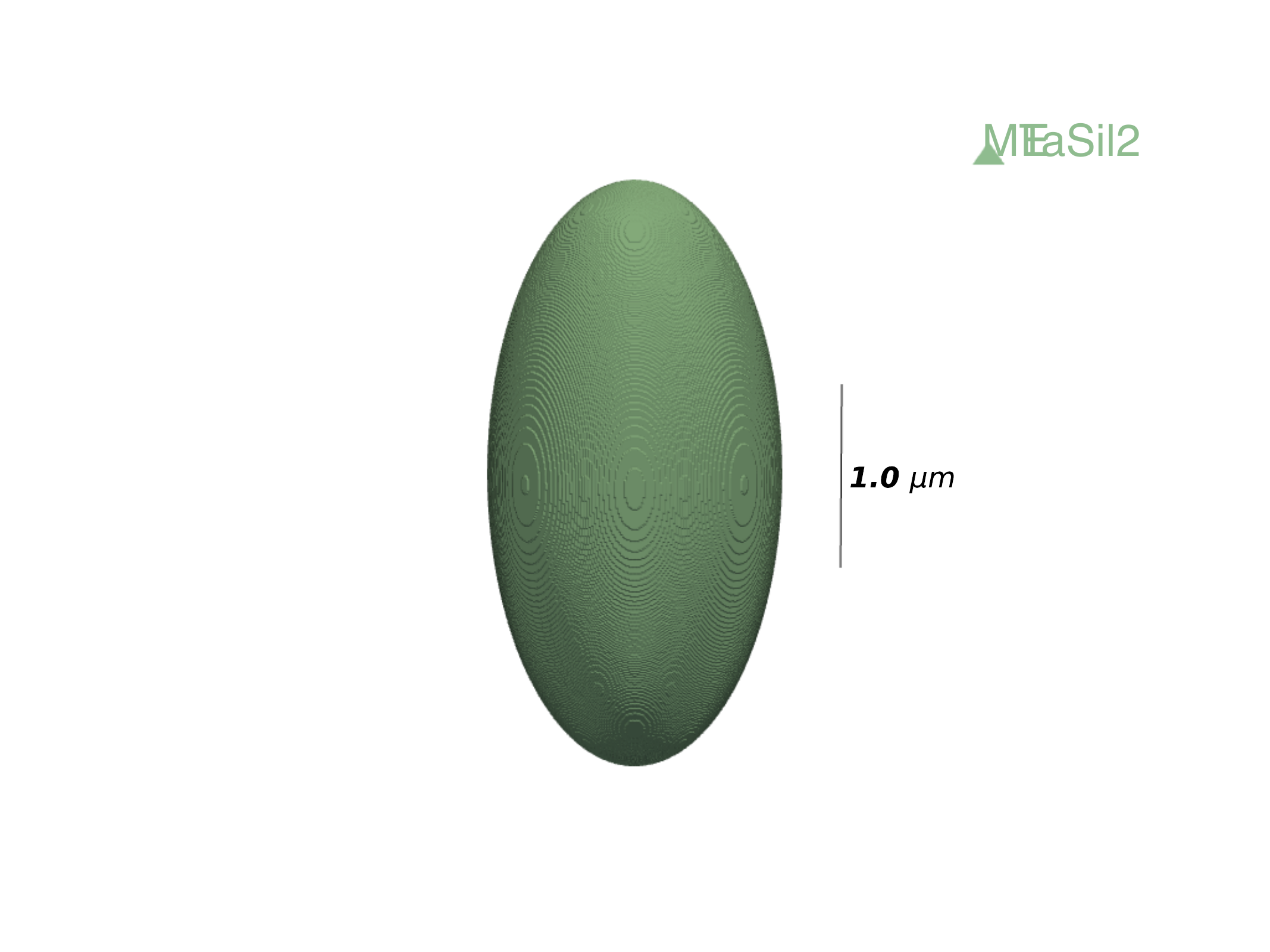}
  \caption{Prol.}
  \label{fig:sfig12}
\end{subfigure}%
\begin{subfigure}{.33\textwidth}
  \centering
  \adjincludegraphics[width=\linewidth,trim={{.08\width} {.13\width} {.08\width} {.13\width}},clip]{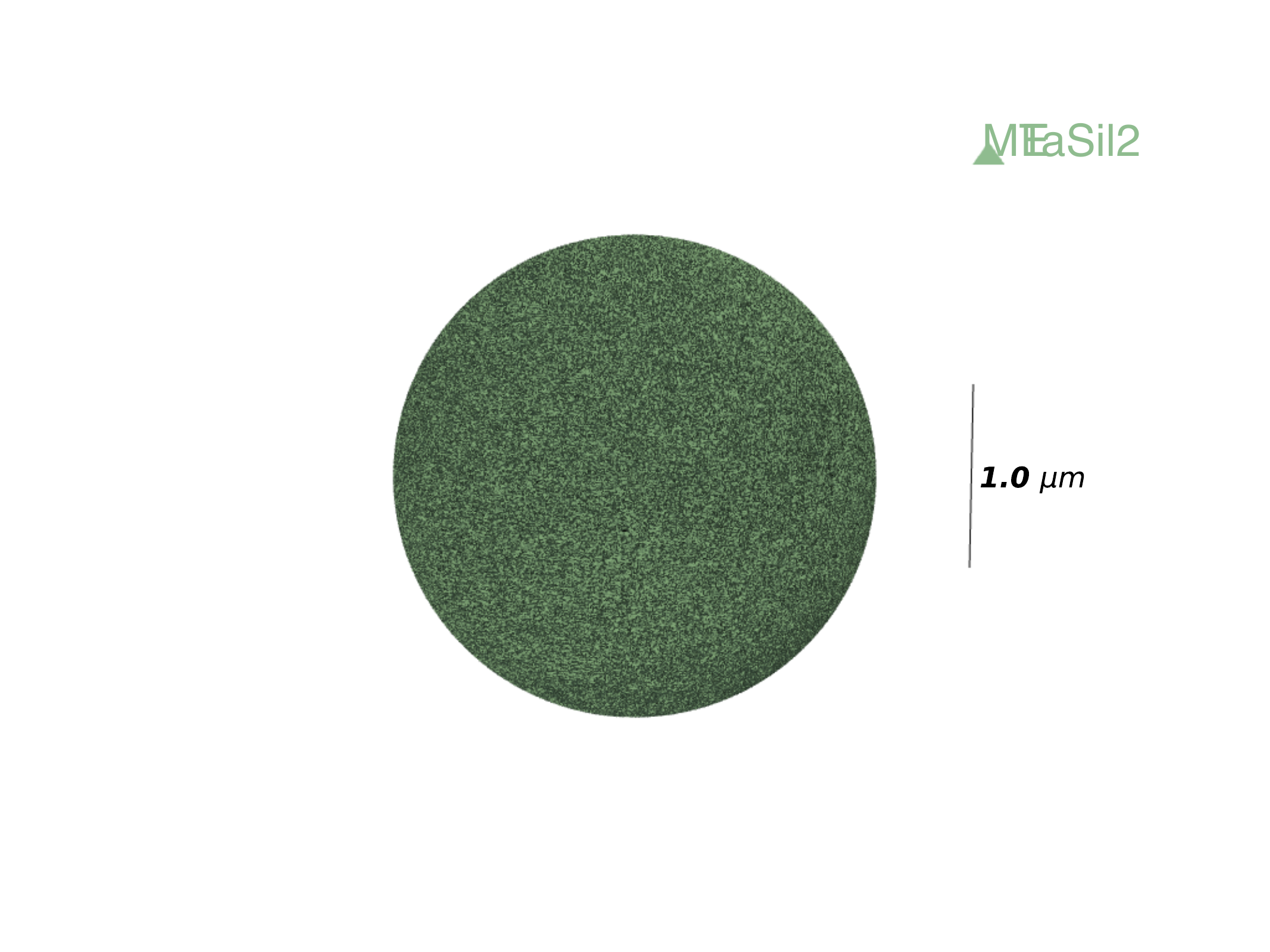}
  \caption{P.Sph.$_{0.54}$}
  \label{fig:sfig13}
\end{subfigure}
\begin{subfigure}{.33\textwidth}
  \centering
  \adjincludegraphics[width=\linewidth,trim={{.08\width} {.13\width} {.08\width} {.13\width}},clip]{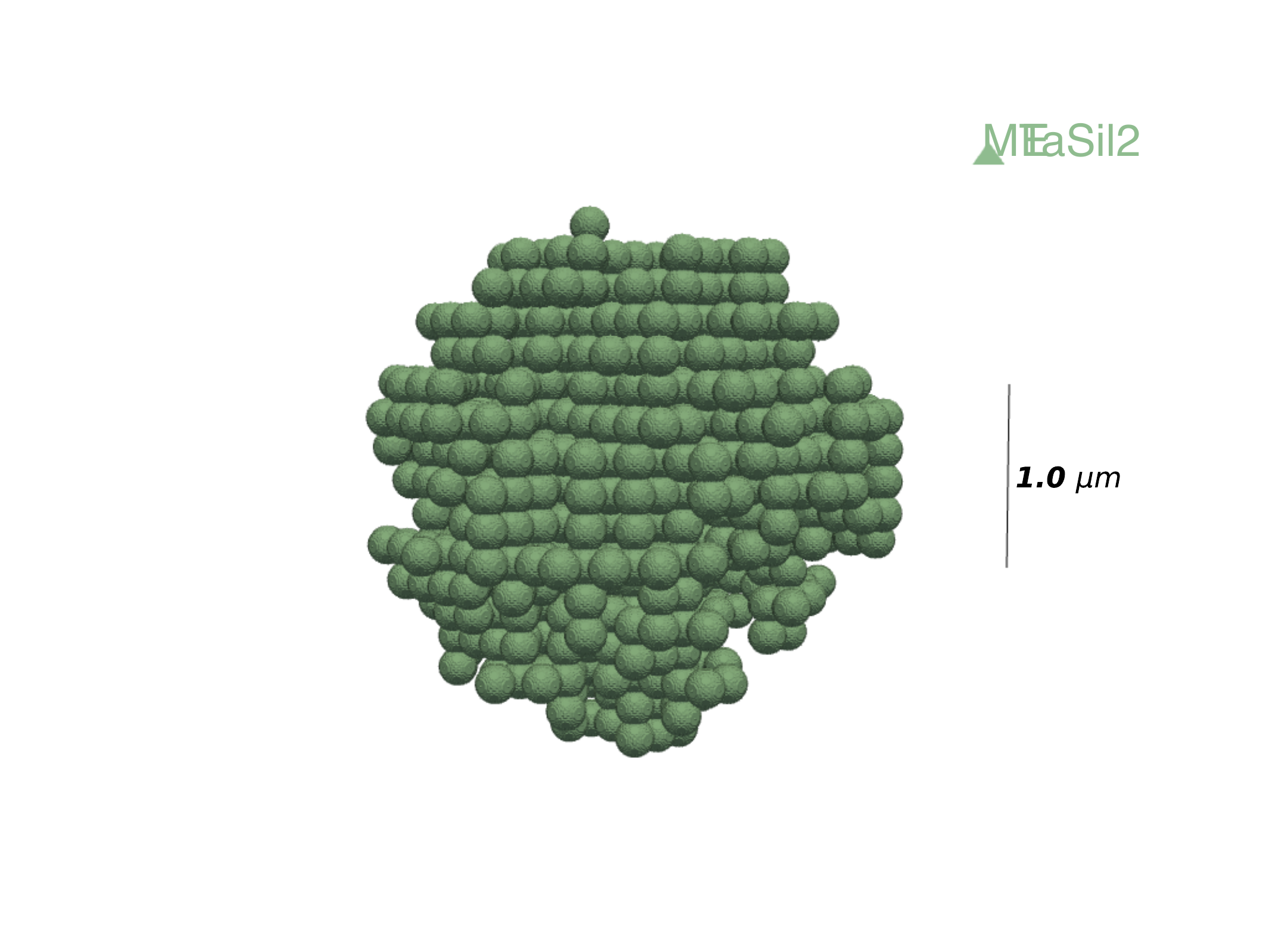}
  \caption{Agg.$^{\qty{100}{\nm}}_{\qty{1}{\um}}$}
  \label{fig:sfig14}
\end{subfigure}
\begin{subfigure}{.33\textwidth}
  \centering
  \adjincludegraphics[width=\linewidth,trim={{.08\width} {.13\width} {0.08\width} {.13\width}},clip]{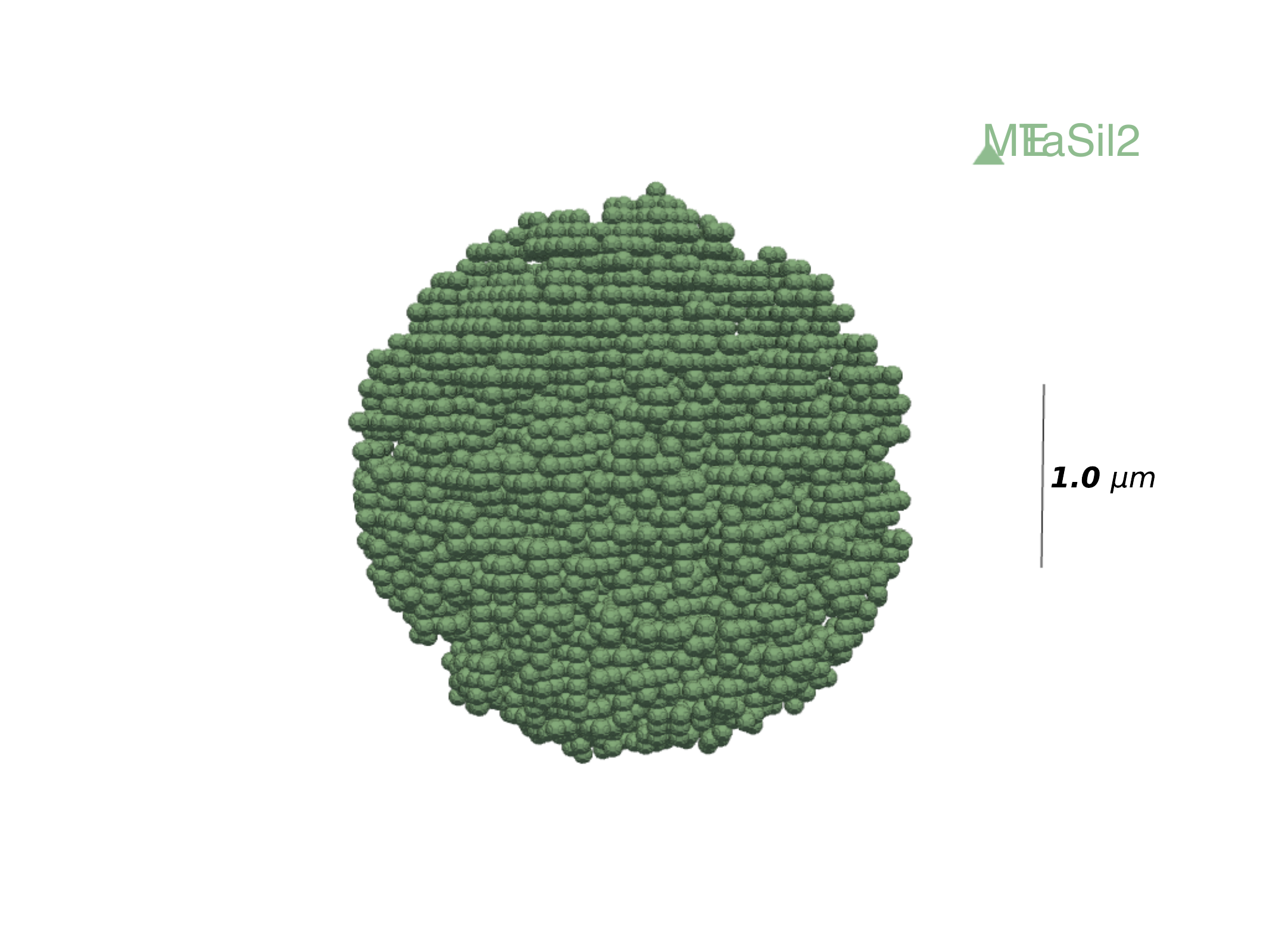}
  \caption{Agg.$^{\qty{50}{\nm}}_{\qty{1}{\um}}$}
  \label{fig:sfig15}
\end{subfigure}
\begin{subfigure}{.33\textwidth}
  \centering
  \adjincludegraphics[width=\linewidth,trim={{.08\width} {.13\width} {.08\width} {.13\width}},clip]{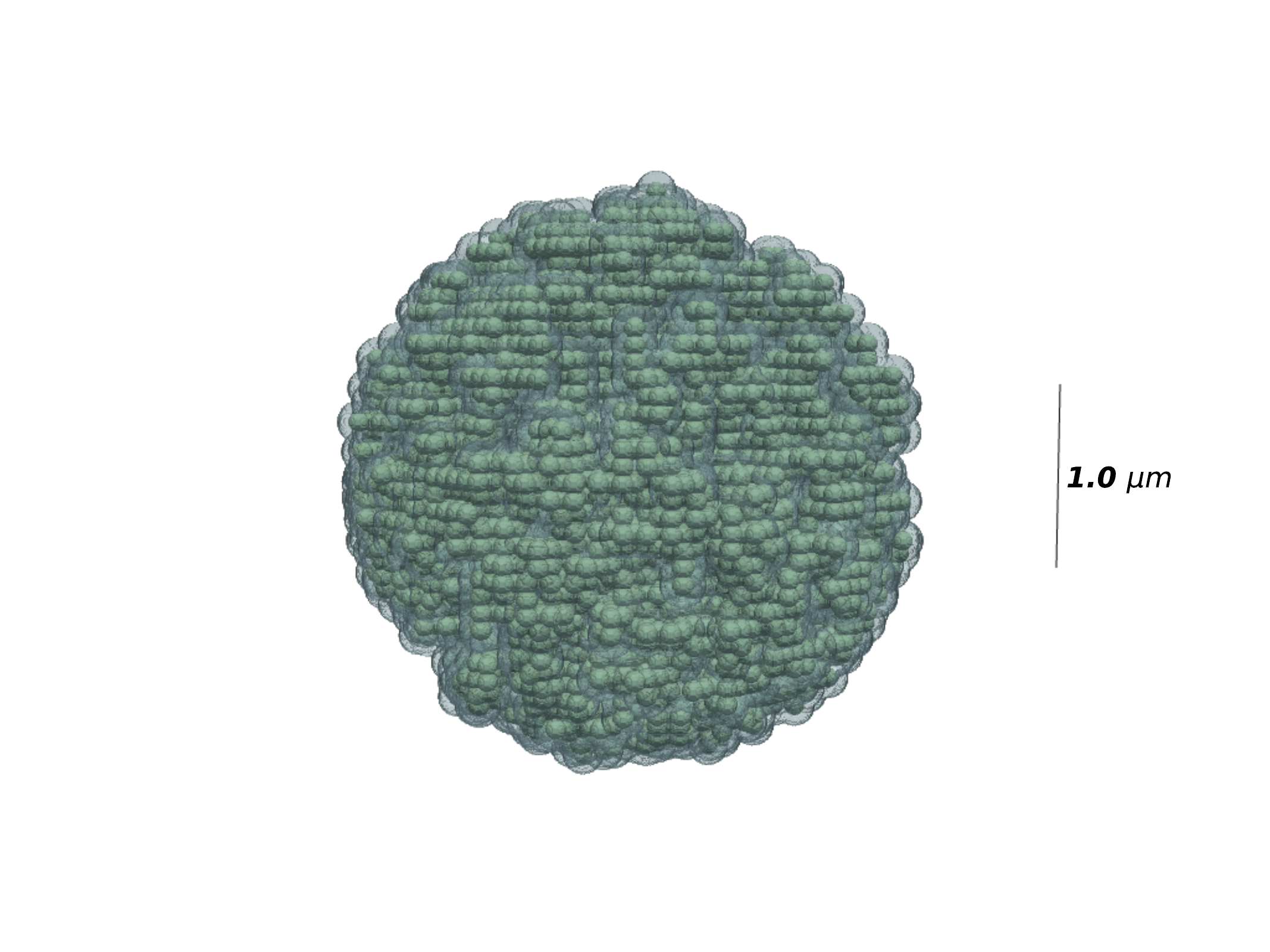}
  \caption{Agg.:ice$^{\qty{100}{\nm}}_{\qty{1.26}{\um}}$}
  \label{fig:sfig16}
\end{subfigure}
\caption{Pilot sample of \qty{1}{\um} dust grains: 3D representations, all grains are shown using the same scale. All grains are describe using dipole sizes of 6.67 nm.}
\label{fig:3Drep}
\end{figure*}

The grain sizes are defined by an "effective radius" $a_{eff}$, which corresponds to the radius of an equal volume sphere:
\begin{equation}    
a_{\eff} = \left( \frac{3V}{4\pi} \right)^{1/3},
\end{equation}
where $V$ is the total grain volume. All the bare grains we model in the pilot sample have an effective radius of \qty{1}{\um}. Note that the one grain to which we add a mantle of water ice has a slightly larger effective radius (\qty{1.26}{\um}).

Another key structural parameter of aggregate dust grains is their intrinsic porosity resulting from the incomplete filling of volume by spherical monomers. Describing such porosity could be achieved by approximating the aggregate grain with an homogeneous porous sphere of equivalent porosity. However, note that whereas it is unambiguous to define porosity in the case of the porous sphere (that is porosity of the bulk material at the microscopic level), defining the porosity of an aggregated grain is more challenging \citep{Jones2011}. Multiple definitions exist \citep{shen_modeling_2008, ossenkopf_dust_1993,kozasa_optical_1992}, here we choose to use the following one:
\begin{equation}
    \label{eq:poros}
    p = 1 - \left(\frac{a_\eff}{\sqrt{5/3}r_\mathrm{g}}\right)^3
\end{equation}
with $r_\mathrm{g}$ the gyration radius of the aggregate, definition that can be found in \cite{kozasa_optical_1992}. While porosity defined in this way does not relate well to the bulk's material's porosity of another object in the case of very fluffy aggregates, \cref{eq:poros} is relevant since the aggregate we consider is rather compact, .

We study six different shapes summarised in \cref{table:grains}, ranging from the simplest sphere to a more complex aggregate (all considered shapes are represented in \cref{fig:3Drep}.):
\begin{itemize}
    \item compact spherical grain;
    \item compact prolate spheroidal grain with an aspect ratio of 2;
    \item porous sphere with a porosity of $p=0.54$;
    \item aggregate composed of 1052 spherical monomers of radius $a_0=\qty{100}{\nm}$, with a fractal dimension $D_f = 2.5$. Using the gyration radius-based definition of porosity \citep{kozasa_optical_1992}, it has a porosity of $p=0.54$;
    \item aggregate composed of 8418 spherical monomers of radius $a_0 = \qty{50}{\nm}$ with a fractal dimension $D_f = 2.5$. Using the gyration radius-based definition of porosity \citep{kozasa_optical_1992}, it has a porosity of $p=0.64$;
    \item as per the previous but coated with a water-ice mantle such that the ice:grain volume ratio is 1:1. Its total effective radius is $a_\eff = 2^{1/3} = \qty{1.26}{\um}$.
\end{itemize}

Multiple methods for constructing aggregates exist between the two extreme cases presented by \citep{kozasa_optical_1992}, which are: BPCA (Ballistic Particle Cluster Aggregates) with a fractal dimension $D_f = 3$ and BCCA (Ballistic Cluster Cluster Aggregates) with $D_f = 2$. Other works also consider varying monomer sizes, such as in \citet{reissl_rotational_2024, jager_radiative_2024}.
In this work, the aggregate grains are constructed following the procedure described in \citet{kohler_aggregate_2011} using single-sized monomers placed on a Cartesian grid according to $D_f$, with a monomer separation of \qty{1.7}{\um}, and choosing an intermediate value of $D_f=2.5$. This choice is justified as we aim to model small grains aggregated from pristine \ac{ism} dust. 
Note that while dust sizes evolve, dust aggregates are likely formed from a much more diverse initial grain population and may require a range of monomer sizes. Studies have already investigated the impact of aggregates composed of monomers of different sizes \citep[e.g.][]{kohler_dust_2012, Liu2015, ysard_optical_2018}, showing that this results in increased scattering and absorption efficiencies across all wavelengths. However, as demonstrated in previous studies, assuming monomers of a uniform size will not change the trends presented in the following, as the results scale with the relative values of the efficiencies.

\begin{table*}[]
\caption{Pilot sample of \qty{1}{\um} dust grains}             
\label{table:grains}      
\centering  
\begin{tabular}{cccccc}
\hline\hline
Acronym                               & Shape                         & {Monomer size  ($a_0, \qty{}{\nm})$}    & Specificity                     &Characteristic radius $a_\mathrm{c}$ (\qty{}{\um})                   & Composition             \\ \hline
\multirow{2}{*}{Sph.}                 & \multirow{2}{*}{Sphere}       & \multirow{2}{*}{1000} & \multirow{2}{*}{/}               & \multirow{2}{*}{1}                 & a-Sil2                   \\ 
                                      &                               &                       &                                              &      & a-Sil7                   \\ \hline
\multirow{2}{*}{Prol.}    & \multirow{2}{*}{Spheroid}     & \multirow{2}{*}{1000} & \multirow{2}{*}{Prolate $\gamma=2$}   & \multirow{2}{*}{1.1}  & a-Sil2                   \\
                                      &                               &                       &                                             &       & a-Sil7                   \\\hline
\multirow{2}{*}{P.Sph.$_{0.54}$}   & \multirow{2}{*}{Sphere}       & \multirow{2}{*}{1000} & \multirow{2}{*}{$p=0.54$\tablefootmark{a}}           & \multirow{2}{*}{1.3}        & a-Sil2                   \\
                                      &                               &                       &                                              &      & a-Sil7                   \\\hline
\multirow{2}{*}{Agg.$^{\qty{100}{\nm}}_{\qty{1}{\um}}$} & \multirow{2}{*}{Aggregate}    & \multirow{2}{*}{100}  & \multirow{2}{*}{$p^\star=0.54$\tablefootmark{b}}                    & \multirow{2}{*}{1.3}               & a-Sil2                   \\
                                      &                               &                       &                                              &      & a-Sil7                   \\\hline
\multirow{2}{*}{Agg.$^{\qty{50}{\nm}}_{\qty{1}{\um}}$}                  & \multirow{2}{*}{Aggregate} & \multirow{2}{*}{50}                    & \multirow{2}{*}{$p^\star=0.64$\tablefootmark{b}}                & \multirow{2}{*}{1.4}                                    & \multirow{2}{*}{a-Sil2}                   \\

                                      &                               &                       &                                                    &                    \\\hline
\multirow{2}{*}{Agg.:ice$^{\qty{50}{\nm}}_{\qty{1.26}{\um}}$}           & \multirow{2}{*}{Aggregate}                     & \multirow{2}{*}{50}                    & \multirow{2}{*}{Ice mantle}                  & \multirow{2}{*}{1.4}                        & \multirow{2}{*}{a-Sil2 + $\mathrm{H_2O}$} \\

                                      &                               &                       &                                                    &                    \\ \hline
\end{tabular}
\tablefoot{
 All grains have an equivalent radius of $a_\eff = \qty{1}{\um}$, except for the iced aggregate, that consists of its non-iced equivalent covered by the same volume of ice: its equivalent radius is thus $a_\eff = \qty{1.26}{\um}$. When not specified, the aspect ratio is by default 1 and the porosity is by default 0. Characteristic radius $a_\mathrm{c}$ is defined as $\sqrt{5/3} r_g$ and represents the overall radius of the grain.\\
\tablefoottext{a}{Porosity of the bulk material}
\tablefoottext{b}{Porosity of the aggregate \citep{kozasa_optical_1992}}

}
\end{table*}

\subsection{Discrete dipole approximation}
\label{subsec:dda}

Multiple options are available to derive the optical constants of dust grains. For instance, Mie theory \citep{mie_beitrage_1908} can provide an analytical solution to the scattering of a spherical particle. However, when it comes to spheroids or irregular shapes such as aggregates, numerical methods have to be introduced. In this study, we opted for the use of the \ac{dda} with the code ADDA \citep[Amsterdam Discrete Dipole Approximation, ][]{yurkin_discrete-dipole-approximation_2011}.
The principle of \ac{dda} is to discretise entirely the grain into small cubes ("dipoles"), and give each dipole a complex refractive index. 

A restriction that rises with the \ac{dda} method is the size of the dipoles. On the one hand, dipoles must be small enough so:
\begin{itemize}
    \item The discretised shape of the grain is geometrically correct and no unwanted surface effects arise.
    \item If we model a mantle, a minimum of 3 dipoles in the mantle thickness is required for a good approximation \citep{kohler_dust_2015}.
    \item The \ac{dda} method is valid under the condition $|m|kd < 0.5$, with $d$ the dipole size, $k = 2\pi / \lambda$ the wave-number and $|m|$ the module of the complex refractive index. 
\end{itemize}
On the other hand, for a given $a_\eff$, reducing the size of the dipoles creates an increase of the dipole number by a power of 3. The code complexity is linear with the dipole number, so using dipoles too small becomes rapidly computationally prohibitive. 

Using THEMIS 2.0 with \ac{emt} permits to relax the second constraint, so the main constraint is to ensure the $|m|kd < 0.5$ criterion.

The grains finally consist of several millions of dipoles so huge computational resources are required. Thankfully, ADDA single calculations can be parallelised using MPI\footnote{Message Passing Interface.}, so we are able to compute efficient and realistic calculations, thanks to the resources of the GENCI at TGCC\footnote{GENCI stands for "Grand Equipement de Calcul Intensif", a high-performance computing facility hosted by the TGCC ("Tr\`{e}s Grand Centre de Calcul du CEA"), an infrastructure able to host petascale supercomputers: \url{https://www.genci.fr/en}.}. 

ADDA computes the interaction of an incoming photon with the dipoles constituting the modelled grain. This calculation can be done at any given wavelength, thus requiring to perform multiple computations to build the spectral dependency of the grain optical properties. In this study, we carry out calculations for 40 wavelengths, logarithmically distributed from \qty{50}{\nm} to \qty{5}{\mm}. For better characterisation of the features, the wavelength sampling has been refined around \qty{3.1}{\um} and \qty{200}{\um} for the \aggice grain, totalling 60 wavelengths for this grain.

For non-spherical particles, it is necessary to also take into account the direction of the incoming photon with respect to the grain surface. ADDA allows to mimic the spatial distribution of incoming photon direction by rotating the discretized grain along its 3 Euler angles ($\alpha$, $\beta$, $\gamma$, see Appendix~\ref{sec:Angles}). For each given orientation, ADDA computes the extinction $Q_\ext$, scattering $Q_\sca$, and absorption $Q_\abs$ efficiencies with $Q_\ext = Q_\abs + Q_\sca$. ADDA also computes the Mueller matrix $\mathbf{S}$ \citep{mueller_memorandum_1943,bohren_absorption_1983}
which encapsulates the mathematical description of how light is altered by an optical material describing both intensity (irradiance) and polarisation changes, including a reduction of the total polarisation. 

Moreover, not only the direction of the incoming photon must be considered, but also the direction of the scattered photons. Thus, ADDA allows us to compute angle-dependent values of the Mueller matrix, for a range of scattering angles $(\theta, \varphi)$. 
Note that the first Euler angle $\alpha$ is perfectly redundant with the azimuthal scattering angle $\varphi$, so it does not need to be sampled. For the aggregates and the spheroids, we compute respectively 24 and 19 \footnote{Taking into account the symmetries of the spheroid.} combinations of the $(\beta,\gamma)$ angles, which permit orientational-averaging.

\section{Results}
\label{sec:results}

From the quantities computed with ADDA ($Q_\abs$,$Q_\sca$, $Q_\ext$, $\mathbf{S}$), the optical properties of all the grains with the necessary number of dipoles and appropriate orientational-averaging, we derive several different quantities to analyse the emissivity, scattering and polarisation of the grains. We present these results in this section.

\subsection{Emissivity of the modelled dust particles}
\label{subsec:emmissivity}

\begin{figure*}[h]
  \centering
\begin{subfigure}{.5\linewidth}
  \centering
  \subfigimg[width=\linewidth,trim={0 {.5\height} {.5\width} 0},clip]{pos=ur, hsep = 0.5cm, vsep =1cm}{\small (a)}{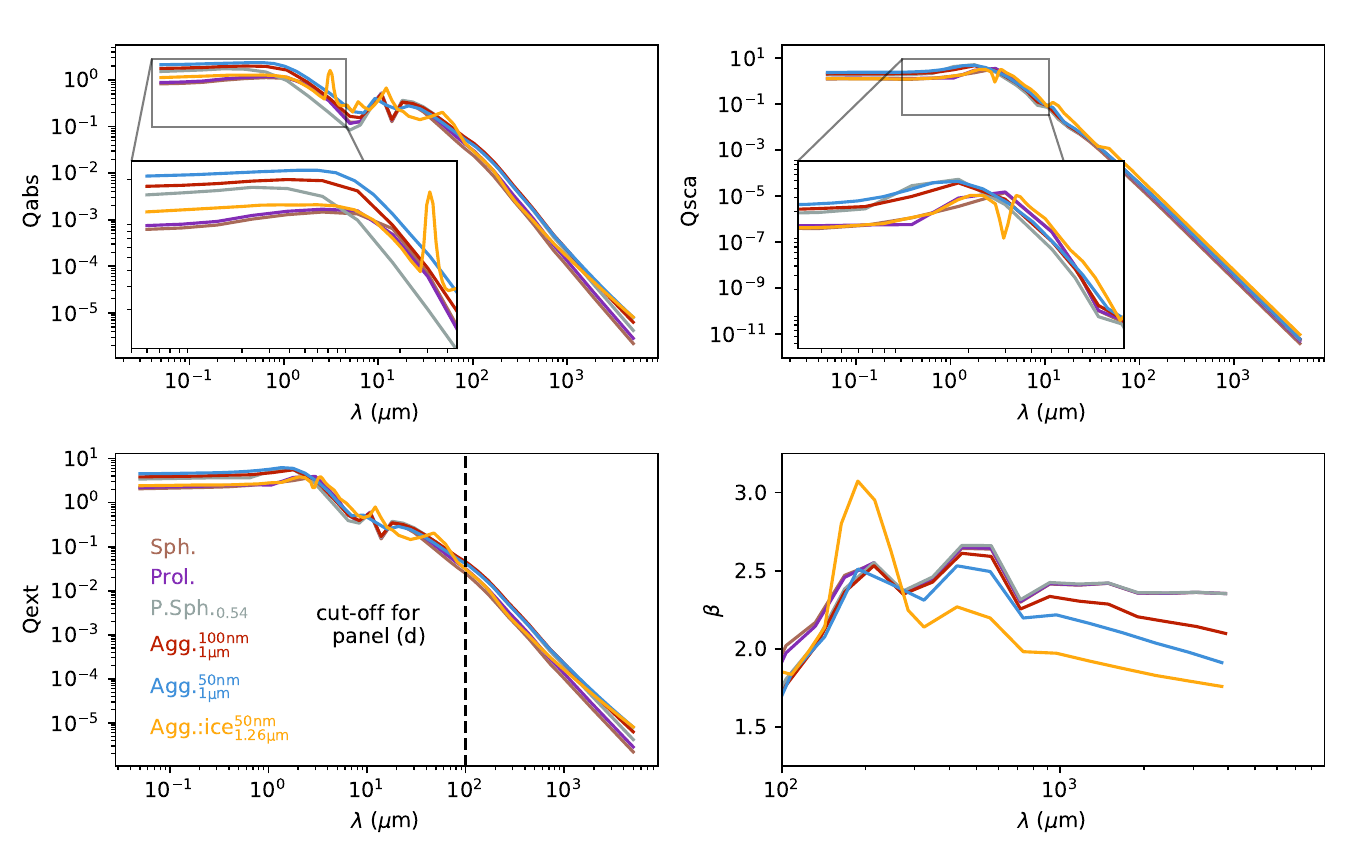} 
  \phantomsubcaption
  \label{fig:sfigQabs}
\end{subfigure}%
\begin{subfigure}{.5\linewidth}
  \centering
  \subfigimg[width=\linewidth,trim={{.5\width} {.5\height} 0 0},clip]{pos=ur, hsep = 0.5cm, vsep =1cm}{\small (b)}{figures/Q.pdf}
  %\caption{}
  \phantomsubcaption
  \label{fig:sfigQsca}
\end{subfigure}%
\newline
\begin{subfigure}{.495\linewidth}
  \centering
  \subfigimg[width=\linewidth,trim={0 0 {.5\width} {.5\height}},clip]{pos=ur, hsep = 0.5cm, vsep =0.85cm}{\small (c)}{figures/Q.pdf}
  % \caption{}
  \phantomsubcaption
  \label{fig:sfigQext}
\end{subfigure}
\begin{subfigure}{.495\linewidth}
  \centering
  \subfigimg[width=\linewidth,trim={{.5\width} 0 0 {.5\height}},clip]{pos=ur, hsep = 0.5cm, vsep =0.85cm}{\small (d)}{figures/Q.pdf}
  %\caption{}
  \phantomsubcaption
  \label{fig:sfigQbeta}
\end{subfigure}
  \caption{Absorption (a), scattering (b), extinction (c) efficiencies and emissivity index (d) for the 6 grains of the pilot sample. We remind that all grains have an effective radius of \qty{1}{\um} for their silicate core and carbon mantle part, while the iced aggregate has a total $a_\eff = \qty{1.26}{\um}$. Panel (d): the emissivity index $\beta$ is computed as the slope of the extinction efficiency, and is shown over a restricted wavelength range, focusing on the sub-mm and \acs{mm}.}
  \label{fig:Q}
\end{figure*}

The emissivity of a dust particle is given by its absorption efficiency $Q_\abs$: it is directly proportional to the absorption cross-section $C_\abs$ and to the probability for a photon to get absorbed by the grain. In \cref{fig:sfigQabs}, we compare the $Q_{abs}$ of the 6 different \qty{1}{\um} a-Sil grains described in \cref{subsec:grainshape}.

The emissivities for all the grains follow the expected common general trend: steady emissivity in the visible at $<\qty{1}{\um}$, a second regime at infrared wavelengths where the emissivity shows spectral features, and finally a power-law decrease at long wavelengths. At short wavelengths, the different grain models produce achromatic but distinguishable emissivity values, all within an order of magnitude, around 1. Such a difference between grains is expected as the grain structure regulates its absorption capability. In the near- and mid-infrared, the emissivity of all grains shows a structured behaviour. Note the absorption features in the \ac{mir}: we see the silicate absorption bands at \qty{9.7} and \qty{18}{\um} and the water absorption bands at \qty{3.1}, \qty{12} and \qty{45}{\um}. In the third regime, where the wavelength $\lambda$ is large compared to the grain size $a_\eff$, we find that the different grain models produce similar but not identical decreases in emissivity with wavelength. The "power-law" curves do not overlay, and deviate from one grain to another. The slopes of these curves and how they vary will be discussed further in the next section.

\subsection{Emissivity index}
\label{subsec:index}

The emissivity index $\beta$ is the spectral logarithmic slope of the extinction efficiency $Q_{\ext}$ in the \ac{fir} to \ac{mm} range. This also equals the slope of $Q_{\abs}$ in the \ac{mm} range, as $Q_{\ext} = Q_{\abs}+Q_{\sca}$ and $Q_{\sca}$ represents a negligible portion of $Q_{\ext}$ ($<10^{-3}\%$) at these wavelengths. We compute the local spectral slope $\beta(\lambda)$ as follows: 
\begin{equation}
    \beta(\lambda) = -\frac{\mathrm{d}log(Q_\ext)}{\mathrm{d}log(\lambda)}
\end{equation}

The wavelength range of interest for $\beta$ corresponds to the third regime described in \cref{subsec:emmissivity} (\qty{100}{\um} $< \lambda <$ \qty{4}{mm}). In this range the wavelengths are sampled on 17 points providing 16 $\beta(\lambda)$ values, shown in \cref{fig:sfigQbeta}. For a fixed wavelength value, we observe different values of $\beta$ from grain to grain, which is influenced not only by the aggregate nature of the grain but also by the substructure of the aggregates themselves. A difference of up to $20\%$ at mm wavelengths can be observed, depending only on the bare shape of the grain. The Agg.:ice$^{\qty{50}{\nm}}_{\qty{1.26}{\um}}$ stands out even more from the non-iced grains.
Note that, contrary to a single power law, the spectral indices vary with wavelength. There is structure in the $\beta(\lambda)$ curve for all grains, implying slope breaks even at wavelengths longer than \qty{1}{\mm}
The emissivity indices range from 1.75 to 2.4 in the \ac{mm} range, and from 2 to 3 in the \ac{sub-mm} range. Even if the $\beta$ overall stabilises in the \ac{mm} range, we note that it remains wavelength (or frequency) dependent.
These variations are entirely due to the choice of optical properties for the silicates we use, which come from laboratory measurements down to the millimetre. The optical properties of amorphous solids are known to vary at long wavelengths (see for instance \citet{Mennella1998} for amorphous carbons and \citet{Coupeaud2011} for amorphous silicates). These variations are explained by the disordered structure of amorphous solids on a small scale as described in \citet{meny_far-infrared_2007}. Previous studies mainly used astrosilicate-type optical properties \citep{draine_optical_1984}, which were extrapolated to long wavelengths with a constant spectral index, as in the case of crystalline solids. The consequence of this choice was therefore a constant spectral index in the Rayleigh domain \citep[e.g.][]{draine_submillimeter_2006, kataoka_opacity_2014}.

\subsection{Scattering}
\label{subsec:scattering}

\begin{figure}[h]
  \centering
  \includegraphics[width=\linewidth,clip]{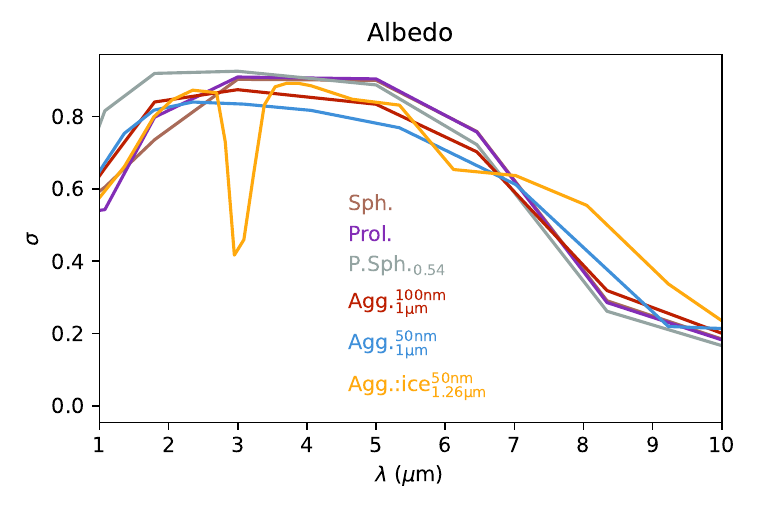}
  \caption{Scattering albedo in the \ac{mir} for the 6 types of silicate grains. All grains have an effective radius of \qty{1}{\um} for their refractive part (the iced aggregate has a total $a_\eff = \qty{1.26}{\um}$).}
  \label{fig:albedo}
\end{figure}

\begin{figure*}[h]
  \centering
  \includegraphics[width=\linewidth,clip]{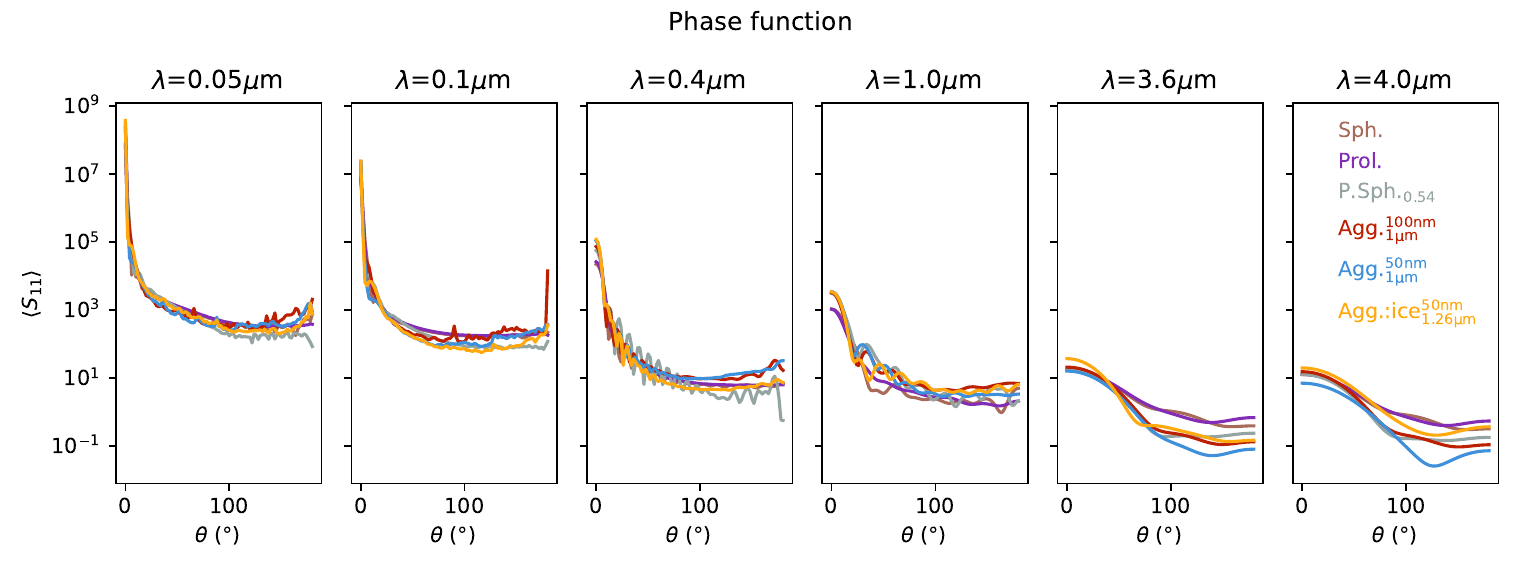}
  \caption{Phase function $\langle S_{11}\rangle$ versus scattering angle. The wavelengths shown were chosen as bracketing the size of the grains $a_\eff$ and monomers $a_0$, as these wavelengths are expected to exhibit the largest scattering efficiencies. All grains have an effective radius of \qty{1}{\um} for their refractive part (the iced aggregate has a total $a_\eff = \qty{1.26}{\um}$).}
  \label{fig:phaseFunc}
\end{figure*}

The integrated scattering efficiencies computed by ADDA, measuring the proportion of scattered light from the total light incoming on a grain, are shown in \cref{fig:sfigQsca}. The three regimes identified in emissivity can also be distinguished, with large monochromatic scattering efficiencies at wavelengths $<\qty{1}{\um}$, large but achromatic values between 1 and \qty{10}{\um}, and negligible scattering ($<10^{-3}$) at longer wavelengths.

To examine the spectral dependency of the scattering efficiencies where they show variation (1-\qty{10}{\um}), we compute the scattering albedo of the grains, defined as $\sigma = Q_\sca / (Q_\sca + Q_\abs)$. 
This quantity is relative to the total amount of light interacting with the grain, and thus easier to use than $Q_\sca$ when comparing the relative importance of scattering from grain to grain.
\cref{fig:albedo} shows that, overall, the scattering albedo of the grains peaks between 0.8 and 0.9 between \qty{1}{\um} and \qty{6}{\um}, and then declines towards zero at longer wavelengths. 
Indeed, at longer wavelengths, the grains are more efficient at absorbing than scattering.
The grain coated with ice shows a dip around \qty{3.1}{\um}, corresponding to absorption in the water band.

The main scattering properties of nonspherical particles  are encapsulated in the Mueller matrix elements $S_{ij}$ \citep[see][]{Mishchenko_light_2000}, which can be used to compute the phase function $\langle S_{11}\rangle$, or the \ac{dlp} $- \langle S_{12}\rangle/\langle S_{11}\rangle$. 
The Mueller matrix is computed by ADDA for each scattering angle $(\theta, \varphi)$, and we repeat the computation for each grain orientation $(\beta, \gamma)$\footnote{See \cref{fig:angles} for visualisation of the grain orientation and scattering direction.}. 
We average $S_{11}$ over $(\beta, \gamma,\varphi)$ to infer $\langle S_{11}(\theta)\rangle$ and show this phase function for the six modelled grains in \cref{fig:phaseFunc}, plotted at different wavelengths of interest. These wavelengths are chosen as bracketing the size of the grains $a_\eff$
and monomers $a_0$, as these wavelengths are expected to exhibit the largest scattering efficiencies\footnote{These wavelengths do not fall right on the wavelength initially sampled in the\ac{dda} computations of the Mueller Matrix, so a simple linear interpolation is performed here.}.

At first sight, all grains show similar phase functions, with no significant departure from the mean trend. 
As with albedo, the scattering efficiency decreases with wavelength. Furthermore, as wavelength increases, the forth- to back-scattering ratio decreases, as can be seen from the flattening of the phase function (from $10^6$ to $10^2$ at respectively $\lambda = 0.05$ and \qty{4}{\um}).
Note that the spheres and aggregate with larger monomers scatter more at longer wavelengths, as expected.

\subsection{Polarisation}
\label{subsec:polar}

\begin{figure}[h]
  \centering
  \includegraphics[width=\linewidth,clip]{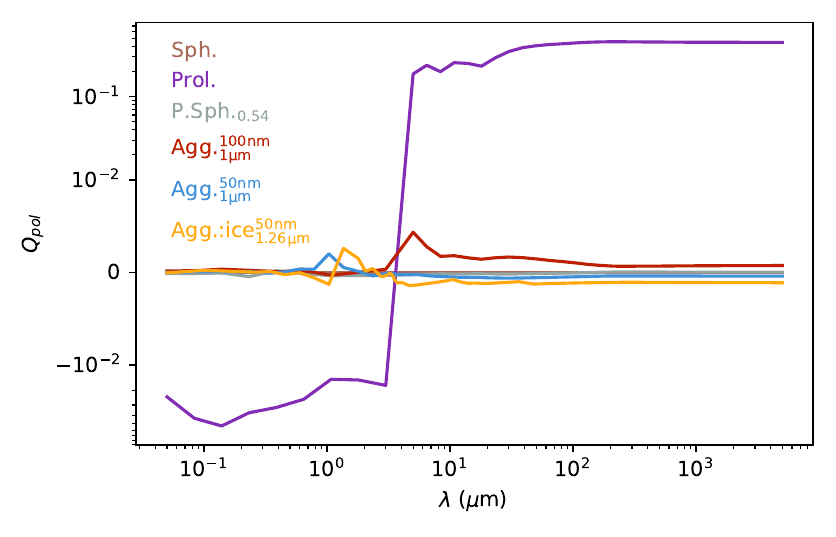}
  \caption{Polarisation efficiency for the 6 types of grains. The y-axis scale is in sym-log, meaning that the values between $-10^{-2}$ and $10^{-2}$ are plotted on a linear scale to allow the 0 crossing, and the remaining of the axis is in log scale.}
  \label{fig:Qpol}
\end{figure}

Starting from the work of \citet{guillet_dust_2018}, we define the normalised direct polarisation induced by a grain (in emission) as $Q_\mathrm{pol}$:
\begin{equation}
    Q_\mathrm{pol} = \frac{1}{Q_\abs}\frac{Q_{\abs,x}-Q_{\abs,y}}{2} =  \frac{Q_{\abs,x}-Q_{\abs,y}}{Q_{\abs,x}+Q_{\abs,y}},
\end{equation}
 $Q_{\abs,x}$ and $Q_{\abs,y}$ being the absorption efficiencies along the two perpendicular directions of alignment $x$ and $y$.
The polarisation efficiencies $Q_\mathrm{pol}$ are shown in \cref{fig:Qpol}. For each grain, the absolute polarisation efficiency peaks and then stabilises around \qty{10}{\um}, while below this threshold $Q_\mathrm{pol}$ fluctuates, mostly for non-spherical grains, with an amplitude up to \qty{6e-2}{}. Overall, we observe significant polarisation only for the Prolate spheroid grain, with absolute values ranging from $10^{-2}$ to 0.4 in the \ac{fir} and mm, while all other grains show values of $10^{-3}$ and below in mm, i.e. no significant polarisation.

\begin{figure*}[h]
  \centering
  \includegraphics[width=\linewidth,clip]{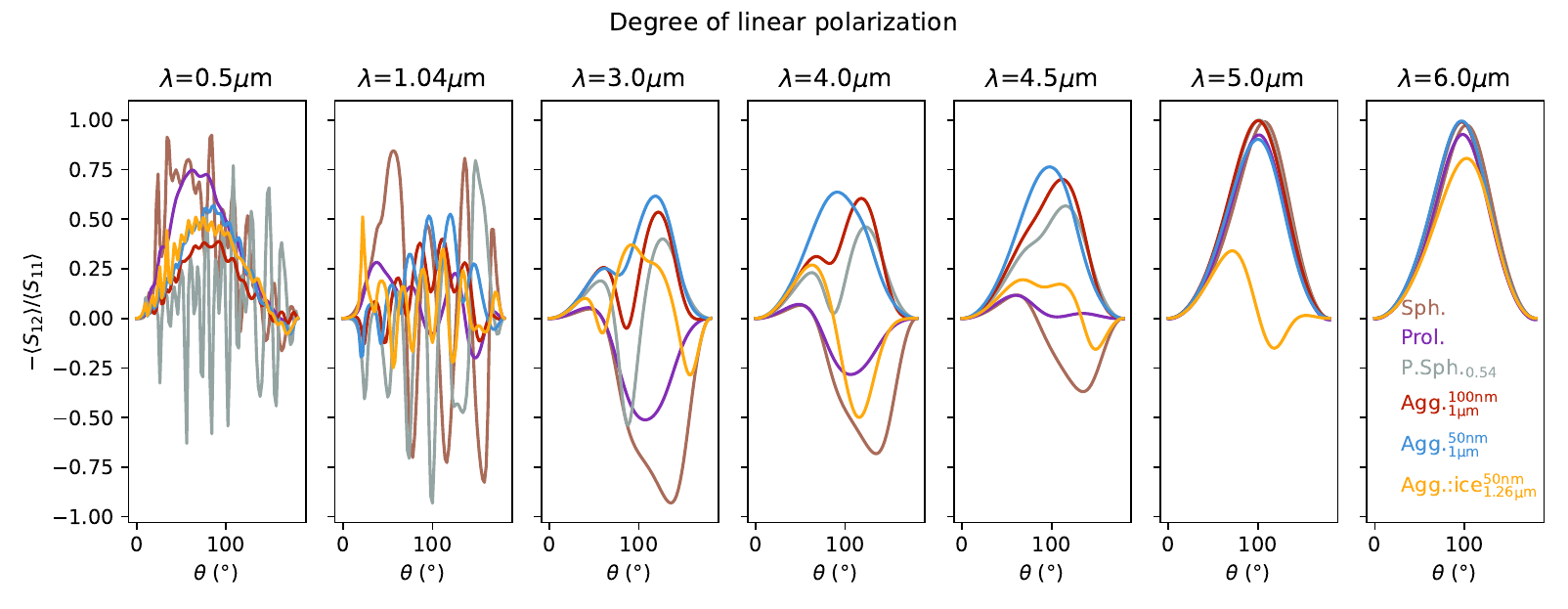}
  \caption{Degree of linear polarisation $- \frac{\langle S_{12}\rangle}{\langle S_{11}\rangle}$ versus scattering angle. The plots are given at different wavelengths of interest depending on the size of the grains $a_\eff$ and monomers $a_0$. All grains have an effective radius of \qty{1}{\um} for their refractive part (the iced aggregate has a total $a_\eff = \qty{1.26}{\um}$).}
  \label{fig:degreepolar}
\end{figure*}

From the Mueller matrix elements, we measure the scattering polarisation by computing the Degree of Linear Polarisation \citep[DLP, see][p.157]{bohren_absorption_1983}:
\begin{equation}
    P = - \frac{\langle S_{12}\rangle}{\langle S_{11}\rangle}.
\end{equation}
In \cref{fig:degreepolar}, we plot $P$ versus the scattering angle $\theta$ for a selection of wavelengths from 0.5 to \qty{6}{\um}. The values are interpolated at the selected wavelengths as in \cref{subsec:scattering}.

At long wavelengths, the degree of linear polarisation of all grains tends to behave in the same way, with a bell-shaped curve and a peak in polarisation for a scattering angle of $\theta = \qty{90}{\degree}$. This corresponds to the Rayleigh scattering \citep{rayleigh_light_1871, bohren_absorption_1983} that we expect when the wavelengths become large compared to the particle size. At smaller wavelengths, the degree of linear polarisation becomes more erratic for all grains as oscillations appear and differences between grains arise. %

\section{Discussion}
\label{sec:discussion}

The optical properties of the six modelled dust grains, presented in \cref{sec:results}, show significant differences from grain to grain. In this section, we discuss the possible origins of these differences, focusing on the \ac{ir} and \ac{mm} ranges where most observational studies use the dust emission either to characterize dust properties or trace indirectly other quantities (gas column densities, masses, magnetic fields, etc.).

We emphasise that this study focuses on the optical properties of one grain at a time, and is therefore not strictly comparable with dust properties obtained from observational probes.
For instance, the cross section $C_\ext = \pi a_\eff^2 Q_\ext $ of a single grain of size $a_\eff$ 
cannot be compared directly to any line-of-sight integrated quantity, as the latter would result from the combined effects of a population of different grains with a size distribution $n(a_\eff) $:
\begin{align}
    \centering
    \langle C_\ext(\nu) \rangle &= \left. \int_{a_\min}^{a_\max} C_{\ext}(a_\eff,\nu) n(a_\eff) \d a_\eff \middle/ \int_{a_\min}^{a_\max} n(a_\eff) \d a_\eff \right.
\end{align}
with no immediate relationship between $\langle C_\ext \rangle$ and the single grain properties.

\subsection{Validity of approximation for dust structure and porosity in the \ac{ir} domain}
\label{subsec:ir}

The first thing to note is the difference between the silicate compact grains and the silicate aggregate grains in terms of emissivity at short wavelengths (\cref{fig:sfigQabs}). 
Considering porosity first, all porous silicate grains (aggregates and porous sphere) show an increased emissivity $Q_\abs$ compared to the compact grains in the $<\qty{1}{\um}$ range, with an increase between 90\% and 160\%. However, in the case of the iced aggregate (yellow curve), the increased in emissivity is moderated. This result illustrates the need to correctly account for grain porosity in the large grains models, which is a natural consequence of random monomer aggregation.
Secondly, we can look into detail at the structure of the porous grains. If the porous sphere (grey line) has a similar emissivity and thus seems a good approximation ($<10\%$ difference) for the optical properties of the aggregate (red line) in the $<\qty{0.5}{\um}$ range, its absorption efficiency largely deviates in the \ac{nir}. In this latter range $Q_\abs$ is lower by a factor of 2 from the aggregate to the porous sphere. We conclude that for wavelengths close to the mean grain size $\sim \qty{1}{\um}$, expected in a dense environment, the modelling of complex aggregates by porous smooth spheres does not allow a robust interpretation of the data in terms of emission and hence extinction.

Another key feature at infrared wavelengths is the presence of spectral features or bands. Silicates produce an absorption band at $\sim$\qty{10}{\um} (seen as an increase of $Q_\abs$, \cref{fig:sfigQabs}) for all silicate grains considered. We stress that the wavelength sampling, which varies slightly between grains, does not allow to distinguish a band shift between bare grains. However, for \aggice, refined sampling around the water ice features allows greater confidence in the shape and location of these bands. Accounting for the presence of a thin water ice layer alters the apparent position of the silicate feature, shifting it to large wavelengths as the water-ice \qty{12}{\mu}m feature partially merges with the silicate feature. This reinforces the need to consider icy grains when observing dust in dense media in the infrared.

Now looking at the scattering properties, for the albedo (\cref{fig:albedo}), the interval from 1 to \qty{7}{\um} is the range where most of the scattering occurs for each grain, with little to no scattering at longer wavelengths. In this range, most of the light goes into scattering, except at \qty{3.1}{\um} for the \aggice grain due to the strong water ice feature at this wavelength. On the other hand, the bare grains (without ice mantle) have a similar shape of albedo curve, but we can still note that the \agg and \aggdemi grains have a lower albedo, especially in the 3 to \qty{7}{\um} range. As with $Q_\abs$, the smaller the monomer, the more the properties deviate from those of compact grains. While monomer sizes are related to the equivalent porosity of the aggregate, we note that the effect on $Q_\abs$ cannot be captured by porosity effect alone \citep{ysard_optical_2018}.
The absolute scattering, $Q_\sca$ (\cref{fig:sfigQsca}), shows the opposite phenomenon, an increase for the aggregates. Hence the importance of the relative nature of the albedo: the aggregates scatter more than the compact grains in absolute terms at these wavelengths, but they absorb even more, so in proportion their albedo is smaller and depends on the monomer size.
As a result, aggregates with small monomers will be observed to scatter less than aggregates with larger monomers, and both these grains will also show lower scattering albedos than the unstructured spheroidal grains. However, the differences in scattering albedo for the grains considered in our sample are small ($<20\%$ difference) and probably cannot be probed with the current astronomical instrumentation. Investigating the resulting net scattering albedo of populations of grains on the line of sight is beyond the scope of this paper, but our parametric study of individual grains does not suggest that large variations should be expected due to grain structure.

The phase function shows the same decrease in scattering with increasing wavelength. In \cref{fig:phaseFunc}, we see that all grains have a predominant forward-scattering phenomenon with $\langle S_{11}\rangle$ several orders of magnitude higher at $\theta = \qty{0}{\degree}$ than at any other scattering angle. At longer wavelengths, when $\lambda$ becomes similar to and larger than the size of the grains $a_\eff$, the phase function flattens out. The scattering is less intense and becomes close to isotropic. In these ranges, the flattening is more pronounced for the compact grains. All aggregates \agg, \aggdemi and \aggice, together with the \por grain, still show at least one order of magnitude between forward- and back-scattering. 
This implies that in 3D modelling of radiation transfer in a dense environment, depending on the choice of grain structure, heating by near-IR photons will be more pronounced for porous or structured grains in the inner parts. This also has an impact for the prediction of photons scattered outwards.
While grain growth mechanisms are not expected to produce spherical compact grains as the main outcome, some specific conditions may lead to compaction of dust grains, especially in disk environments when large dust particles fragment \citep{michoulier_compaction_2024}. However, such mechanisms are unlikely to be very efficient for micron-sized grains. Moreover, phase functions from \ac{nir} observations probing the $\sim \qty{1}{\um}$ dust grains at the surface of disks also suggest that the polarised scattered light comes from Mie scattering of compact grains, favouring particle aggregates as opposed to compact spheres \citep{engler_investigating_2019,adam_characterizing_2021}.
Further modelling of complete dust populations of different sizes and shapes, as well as observational constraints on the polarised emission of dust at different wavelengths are required to lift degeneracies in grain elongation and compactness as they evolve from the diffuse ISM to planet-forming disks.

The \acl{dlp} is plotted in \cref{fig:degreepolar}. Once again we see a divergence in behaviour between compact and porous grains. As the wavelength becomes large compared to the particle size, as expected, all grains enter the Rayleigh regime (bell shape): for micron-sized grains, the \ac{dlp} does not contain any information about grain structure or porosity at $\lambda> \qty{5}{\um}$. However, non-compact grains reach this regime earlier than compact grains: \aggdemi already has a Rayleigh-like \ac{dlp} at \qty{4}{\um}, and \agg reaches this behaviour around \qty{4.5}{\um}, while compact grains do not reach it before \qty{5}{\um}. Thus, the spectral variation of the \ac{dlp} can be a good tool to distinguish between grain shapes, should they be compact or aggregated, and could also allow to distinguish between monomer sizes as well \citep[e.g.][]{Halder2018, Halder2021}. The \por grain follows the trend of \agg quite well and is therefore a good approximation of the aggregate for the \ac{dlp} . We note that the \aggice (yellow line) reaches the Rayleigh regime only beyond \qty{6}{\um}, which is due to its larger effective radius $a_\eff = \qty{1.26}{\um}$ (similar values of the dimensionless coefficient $X = 2\pi a_\eff / \lambda$ \citet{tobon_valencia_scattering_2024, tazaki_scattering_2021} of about 1.35 are reached at lambda $\lambda = \qty{6}{\um}$ for the iced aggregate while other aggregate grains reach these values around lambda $\lambda = \qty{4.5}{\um}$). We emphasise the sensitivity of the \ac{dlp} to grain effective radius, which could be a great tool for grain size measurement. However, this sensitivity could be affected in practice when a whole population of dust sizes is present, rather than the individual grains we consider.

\subsection{Effects of approximating the dust porosity and structure for optical properties in the mm domain} 
\label{subsec:mm}

With this pilot sample we are interested in (i) the effect of size on the millimetre emissivity, i.e. do these \qty{1}{\um} grains have a very different \ac{fir}/\ac{mm} $\beta$ from the \qty{0.1}{\um} grains for the diffuse medium given in \citet{ysard_themis_2024} and (ii) the effects of porosity and structure on the millimetre emission, in thermal emission and polarisation, in order to discuss whether it is reasonable to approximate these micron-sized grains by spheres when interpreting the data obtained at millimetre wavelengths. Finally, we recall that these grains form the first sample of aggregates from which hypotheses can be constructed to predict the properties of larger aggregates, whose effective sizes will be closer to \ac{fir}/\ac{mm} wavelengths.

\subsubsection{Optical properties at long wavelengths: aggregates are different from compact grains}
\label{subsec:mm-shape}

For the compact and porous spherical grains, our models predict spectral indices $\beta\sim2.36$ at \qty{3}{\mm}. However, our modelled grains made of aggregated monomers show lower $\beta\sim 1.97-2.15$ for the \aggdemi and \agg respectively. 
Our pilot sample thus shows that the spectral index of the dust emissivity at long wavelengths depends on the grain structure: we find significantly lower dust grain emissivities for the aggregated monomer grains than for the compact grains (see \cref{fig:Q}).
We also note that, in our models of aggregate grains, the smaller the size of the monomer, the lower the emissivity index, given a fixed total volume of material ($a_\eff$ = \qty{1}{\um}). 
Finally, note that approximating the aggregate structure using porosity alone does not allow realistic emissivity indices to be obtained, as the $\beta$ of the porous sphere remains very close to the compact sphere values and does not mimic that of the aggregate.
The physical origins of these long wavelength behaviours are discussed below.

The reduction in the emissivity index of the aggregates with respect to the compact grains has a subtle cause that lies in the \ac{cm} structure of the grain material \citep{ysard_themis_2024} rather than in the purely geometrical modification of the shape. Indeed, all THEMIS 2.0 a-Sil grain models developed for the diffuse medium are coated with a \qty{5}{\nm} mantle of amorphous carbon, irrespective of their size. In the model framework, this mantle results from the accretion of small carbon grains or C atoms from the \ac{ism} onto larger silicate seeds, in the diffuse \ac{ism}. The relative abundances of carbon and silicate material in the global grain population are maintained by mixing pure carbon grains with these silicate-carbon grains. 
Starting from a reservoir of the diffuse \ac{ism} grains, several scenarios can be considered to describe the evolution of dust in the dense medium towards larger grain sizes. 
\\First, the simplest approach is a scaled-up version of the mechanisms at work in the diffuse \ac{ism}, e.g. compact growth of a silicate core that retains the spherical geometry of the grain as it grows, then later becomes coated with the smaller carbon grains: this eventually produces large spherical particles of a-Sil, coated with a thin a-C mantle. For a \qty{1}{\um} spherical grain (such as our compact \sph), $0.7\%$ of the total mass of the grain is in the form of an a-C mantle for a \qty{5}{\nm} mantle depth. 
\\The second, more realistic scenario, which we explore here, is the aggregation of the small \ac{cm} monomers from the diffuse \ac{ism}, which is enhanced as the gas density increases in the dense medium. In such a scenario, a \qty{1}{\um} aggregate made up of \qty{100}{\nm} \ac{cm} monomers, each of which contains $7\%$ of its mass under a-C mantle form, will eventually result in a larger a-C to a-Sil mass ratio than the compact spherical grain of the same size \footnote{Note that the a-C to a-Sil mass ratio of such a \ac{cm} aggregated grain is nearly equivalent to that of the aggregated grain made from pure a-Sil monomers and then coated with a thin a-C mantle.}. 
\citet{ysard_optical_2018} showed that the \ac{mm} emissivity of the THEMIS a-C material is much flatter than the emissivity of a-Sil materials, for grains made entirely of either one or the other. We can deduce that the flattening of the emissivity of the aggregates compared to the compact shapes is likely due to the increased proportion of a-C material relative to the a-Sil material. 
We emphasise that the relative mass of the a-C shell increases as aggregates are built from even smaller monomers. This explains the lower emissivity index of the \aggdemi grain compared to the \agg grain.
Moreover, we find that for a fixed a-C to a-Sil abundance ratio in the grain, both the small grains ($\qty{0.1}{\um}$ representative of the diffuse medium \citep[Table 2]{ysard_themis_2024} and the large aggregate grain \agg have very similar dust emissivity indices $\beta$ at millimetre wavelengths: a change in grain size of one order of magnitude thus, and unsurprisingly, does not affect the optical properties of the grains when at wavelengths much larger than the grain size. To significantly reduce the $\beta$ at millimetre wavelengths, one would therefore need to consider a large fraction of dust aggregates of typical sizes $>\qty{100}{\um}$. The current micron-sized pilot dust grains studied here will provide a good elementary brick as monomers of such very large dust particles.

In summary, the detailed structure of dust grains at the micron scale is not the key to determining their emissivity indices at millimetre wavelengths. However, we stress that the hypothesis made to build up large grains from a population of smaller grains has a significant impact on the expected optical properties at long wavelengths. Indeed, aggregates made of \ac{cm} monomers inherently have larger mass ratios of a-C to a-Sil, which, in turn, will lead to smaller dust emissivity indices $\beta$, following the a-C material behaviour.
Further calculations considering complete dust populations and radiative transfer simulations are needed, but this result is promising in providing an alternative explanation for the low emissivity measured in protostellar environments \citep{galametz_low_2019,cacciapuoti_faust_2023,bouvier_orion_2021}, which can currently only be explained by a large fraction of grains $>\qty{100}{\um}$ \citep[e.g.][]{ysard_grains_2019}, and is at odds with typical timescales from current dust evolution models \citep{ormel_dust_2009,silsbee_dust_2022}. %

In \cref{subsec:polar}, it is noted that the only grain with a significant polarisation efficiency in the far-IR and millimetre range, $Q_\pol >1\%$ is the \prol grain, due to its unique privileged axis, in contrast to the symmetric \sph and \por grains. The aggregates studied, with a high fractal dimension (2.7), are overall spherical and therefore show only very little ($<0.1\%$) intrinsic polarisation. Although it remains small, this polarisation varies with monomer size: grains with larger monomers are less spherical and thus polarise light more as shown by \citet{tazaki_size_2022} under disk conditions. 
Note also that the presence of ice mantles triples the polarisation efficiency $Q_\pol$ (see \cref{sec:Iceaddidion}).
First, we should note that the intrinsic polarisation of a population of prolate grains may not produce any detectable polarised light at far-IR and millimetre wavelengths if the dust particles have no preferred orientation, as the $x$ and $y$ axes would be randomly oriented. The efficiency of polarisation and the efficiency of dust grain alignment are therefore highly degenerate from observational constraints: detailed studies of the alignment physics of complex dust aggregates in magnetised environments, for example, may in the future provide us with a framework allowing to test predictions from different dust shapes.
While theory \citep{weidenschilling_formation_1993,ormel_dust_2007,dominik_physical_2009} predicts that the first stages of dust growth in the pristine solar nebula create porous fractal aggregates, the prevalence of fractal structures in dust aggregates is also supported by experimental work \citep{blum_growth_2008} and by the detection of fractal aggregates in cometary dust \citep{mannel_fractal_2016}.
This initial fractal building may later be lost by collisional compaction of the largest dust particles ($>\qty{}{mm}$ sizes) in high density environments, but dust grains evolving in protostellar environments may retain a relatively high fractal dimension, and therefore elongation, although this is still the subject of ongoing research \citep{tazaki_fractal_2023}.

\subsubsection{Effects of the ice coating}

\label{subsec:mm-ice}

In dense media, protected from the harsh high-energy radiation field at high extinction, low temperatures allow gas to condense on the solid particles. The dust grains are therefore expected to be covered by a mantle of ice, the composition of which will vary according to the local environment. Adding an equal volume of ice to the aggregate grain results in a small difference in effective size, but also marginally modifies the structure of the grain, as the ice is added not only as an outer shell on the aggregate, but also around each monomer. The effects discussed here are a consequence of the properties of water ice rather than these geometrical modifications, as discussed in \cref{sec:Iceaddidion}.
In the case of the iced aggregate, we see in \cref{fig:Q} that the \ac{mm} emissivity index is $\sim 10\% $ lower than that of the non-iced counterpart. While such a small difference may not be captured by the current accuracy of observational measurements of the dust emissivity index, the presence of grains covered by layers of ice at specific locations where the temperature drops, within a given object, may result in a flattening of the emissivity. Thus, not only grain growth and complex structures such as those naturally formed by aggregation of grains (see \cref{subsec:mm-shape}) can explain the overall dust emissivity variations observed in some astrophysical environments. 
Note that, if ice mantles are responsible for low emissivities at mm wavelengths, the spectral feature of silicates at \qty{10}{\um} should be shifted towards longer wavelengths (see \cref{subsec:ir}).

\subsubsection{Effects of silicate composition}
\label{subsec:mm-comp}

\begin{figure*}[h]
  \centering
\begin{subfigure}{.5\linewidth}
  \centering
  \subfigimg[width=\linewidth,trim={0 {.5\height} {.5\width} 0},clip]{pos=ur, hsep = 0.5cm, vsep =0.95cm}{\small (a)}{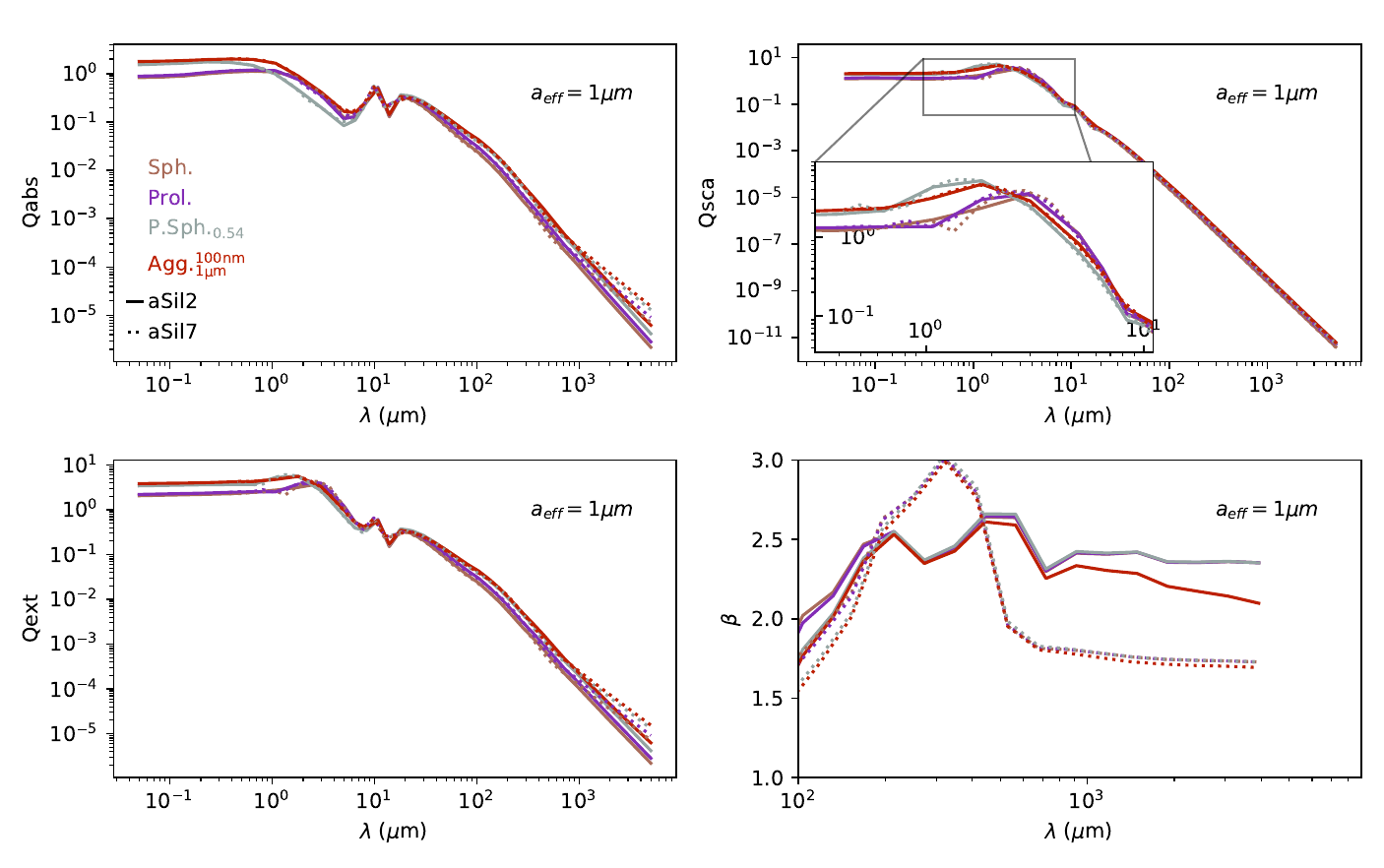}
  \phantomsubcaption
  \label{fig:sfigQcompabs}
\end{subfigure}%
\begin{subfigure}{.5\linewidth}
  \centering
  \subfigimg[width=\linewidth,trim={{.5\width} {.5\height} 0 0},clip]{pos=ur, hsep = 0.5cm, vsep =0.95cm}{\small (b)}{figures/Qcomp.pdf}
  \phantomsubcaption
  \label{fig:sfigQcompsca}
\end{subfigure}%
\newline
\begin{subfigure}{.495\linewidth}
  \centering
  \subfigimg[width=\linewidth,trim={0 0 {.5\width} {.5\height}},clip]{pos=ur, hsep = 0.5cm, vsep =0.8cm}{\small (c)}{figures/Qcomp.pdf}
  \phantomsubcaption
  \label{fig:sfigQcompext}
\end{subfigure}
\begin{subfigure}{.495\linewidth}
  \centering
  \subfigimg[width=\linewidth,trim={{.5\width} 0 0 {.5\height}},clip]{pos=ur, hsep = 0.5cm, vsep =0.8cm}{\small (d)}{figures/Qcomp.pdf}
  \phantomsubcaption
  \label{fig:sfigQcompbeta}
\end{subfigure}
  \caption{Absorption (a), scattering (b), extinction (c) efficiencies and emissivity index (d) for 4 grain structures. The solid lines show the case where the silicate component is represented by the a-Sil2 material described by \citet{ysard_themis_2024} and the dotted lines by the a-Sil7 material.
  All grains have an effective radius of \qty{1}{\um} for their refractive part. The emissivity index is computed as the slope of the extinction efficiency, and is zoomed in the sub-mm and \acs{mm}.}
  \label{fig:Qcomp}
\end{figure*}

\begin{figure}[h]
  \centering
  \includegraphics[width=\linewidth,clip]{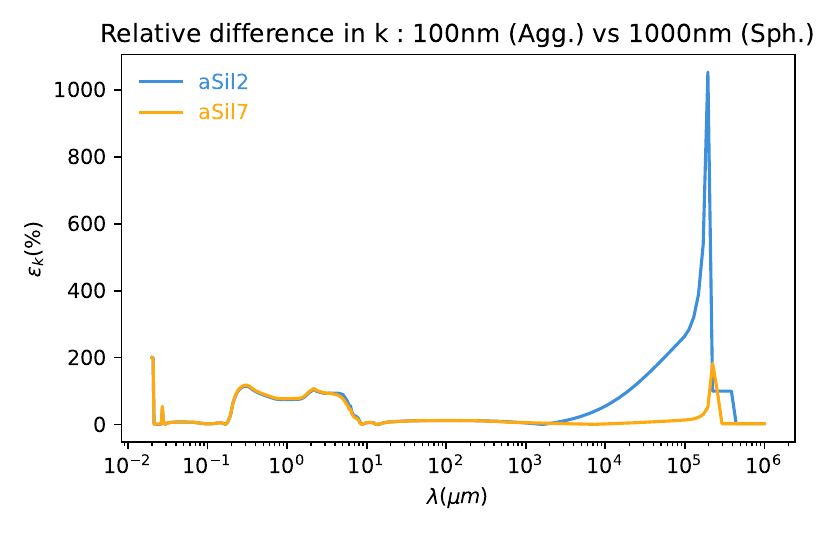}
  \caption{Relative difference between the imaginary part of the refractive index of the \sph (EMT for a \qty{1000}{\nm} sphere) and of the \agg (EMT for a \qty{100}{\nm} sphere). In blue a-Sil2, in orange a-Sil7.}
  \label{fig:EMTcomp}
\end{figure}

Throughout the study we use the a-Sil2 material. A relevant addition to the comparisons made in the previous sections is the comparison of the previous grains (a-Sil2) with grains considering a different silicate composition: a-Sil7, which is built with different a forsterite/enstatite abundance ratio and a different enstatite sample, and therefore shows large deviations of its spectral index $\beta$ \citep[see][Table 2]{ysard_themis_2024}. For the sake of simplicity, in this section we limit the study to 4 grains: \sph, \prol, \por and \agg grains. For the two silicate compositions considered, we plot the $Q\abs$, $Q\sca$, $Q\ext$, and $\beta$ in \cref{fig:Qcomp}. We can see that the emissivity index of the a-Sil7 sphere is significantly lower than that of the sphere made of a-Sil2, meaning that the a-Sil7 silicate material intrinsically has a flatter emissivity in the mm, making it closer to the a-C behaviour. The direct consequence of this is that, despite the associated increase in the mass fraction of carbon material, looking at an aggregate dust grain rather than a compact sphere results in a much less dramatic change in the grain emissivity index for the a-Sil7 material than for a grain of a-Sil2.
As explained in \cref{subsec:opt.const}, we use the EMT approximation, so that the entire a-Sil core / a-C mantle grain structure is encapsulated in one single mock-up material. 
\Cref{fig:EMTcomp} plots the relative difference $\epsilon_k= \frac{k_{\mathrm{sph.}}-k_{\mathrm{agg.}}}{k_{\mathrm{sph.}}}$ of the imaginary refractive index between the sphere EMT and the aggregate EMT, for the two materials we considered, a-Sil2 and a-Sil7.
It shows that since the refractive index of a-Sil7 is already small and close to that of a-C \citep{ysard_themis_2024}, the transition from a sphere to an aggregate adds a-C material but changes only marginally the refractive index for the a-Sil7 aggregate grain, while the effect of adding a-C has much more dramatic consequences on the refractive index of the a-Sil2 aggregate at mm wavelengths. 

Regarding the optical properties of dust grains at mm wavelengths, changes in composition, structure and size appear to have significant effects. However, considering grains of sizes up to $\sim\qty{1}{\um}$, such as those modelled in this study, does not significantly change the $\beta$ at mm wavelengths. Thus, assuming no larger grains are present, different compositions could be another possible explanation for the diversity of emissivity indices observed among astrophysical objects. Even if the stoichiometry of silicates can change depending on local irradiation conditions \citep[e.g.][]{Demyk2001}, it is not clear how composition of silicates could change within a single dust reservoir that only exhibits small variations of the local physical conditions, such as protostellar envelopes. Other considerations of dust processing mechanisms in specific environments and their consequences for observed dust emissivities, should be further investigated.

\section{Conclusions}
\label{sec:concl}

We have studied the optical properties of a pilot sample of six micron-sized dust grains, as analogues of the dust expected to be found in the dense ISM, from the evolution of the small grains shown to be in agreement with diffuse ISM observations  \citep{ysard_themis_2024}. Our goal is to evaluate the validity of approximations made on the grain structure when computing the optical properties of complex grains born from the aggregation of small core-mantle grains. 
With this study, we have focused on exploring the effects of size and grain structure on the resulting optical properties in the IR and mm range.

For all six grains we have investigated emissivity, scattering and polarisation. The main findings of our study are detailed below.
\begin{enumerate}
    \item Up to 20\% differences in emissivity indices, in mm, are found depending on the structure of a grain. Aggregated grains have a significantly lower value of $\beta$ than their spherical or spheroid equivalent for \qty{1}{\um} grains.  
    \item When constructing dust grains, and especially aggregates, it is crucial to question the a-C/a-Sil ratio in the constitution of the grain. Particular care must be taken when constructing a dust population to select the correct total relative abundances. Because of the potential differences in the refractive indices of the components, this ratio directly affects the value of the \ac{mm} emissivity index.
    \item The composition of the silicate part of the grain has a significant impact on the emissivity index values. Changing the stoichiometry of silicate mixtures alone changed the $\beta$ by 25\%, as predicted by the data from \citet{ysard_themis_2024}.
    \item The adjunction of water-ice mantles to dust grains, which may be common in the dense ISM, also considerably affects the value of the emissivity index. We found that the spectral indices are almost 10\% lower when the grain is covered by an ice mantle, which accounts for half of the total volume.
    \item Lastly, polarisation is highly dependent on the shape of the grains. Asymmetry, and especially high aspect ratio, induces more polarisation. This means that spheroid grains, oblates or prolates, are good polarisers. Aggregates can also become polarisers when they are fluffy or contain only a few monomers.
\end{enumerate}

These results prove the importance of a good structure representation in the dust models and the need for comprehensive dense medium grains. 
These grains are the first sample of aggregates from which hypotheses can be made to predict the properties of larger aggregates: a complete grain model, including grains of larger sizes and different structures, will be developed in the future and presented in forthcoming papers.
This pilot study will be the basis for performing detailed physical models including dust radiative transfer simulations, and ultimately interpret observations of dust.

\begin{acknowledgements}
This project was provided with computing HPC and storage resources by GENCI at TGCC thanks to the grant 2024-A0170415717 on the supercomputer Joliot Curie's ROME partition.
Color sequences in plots of this project are designed to be accessible thanks to \citet{petroff_accessible_2021}.
This work was made possible thanks to the support from the European Research Council (ERC) under the European Union’s Horizon 2020 research and innovation programme (Grant agreement No. 101098309 - PEBBLES).
\end{acknowledgements}

\bibliographystyle{aa} 
%\bibliography{bibliography} 

%-------------------------------------------------------------------

\begin{appendix}

\section{Computing time}
\label{sec:CPUtime}

\begin{figure}[h]
  \centering
  \includegraphics[width=\linewidth,clip]{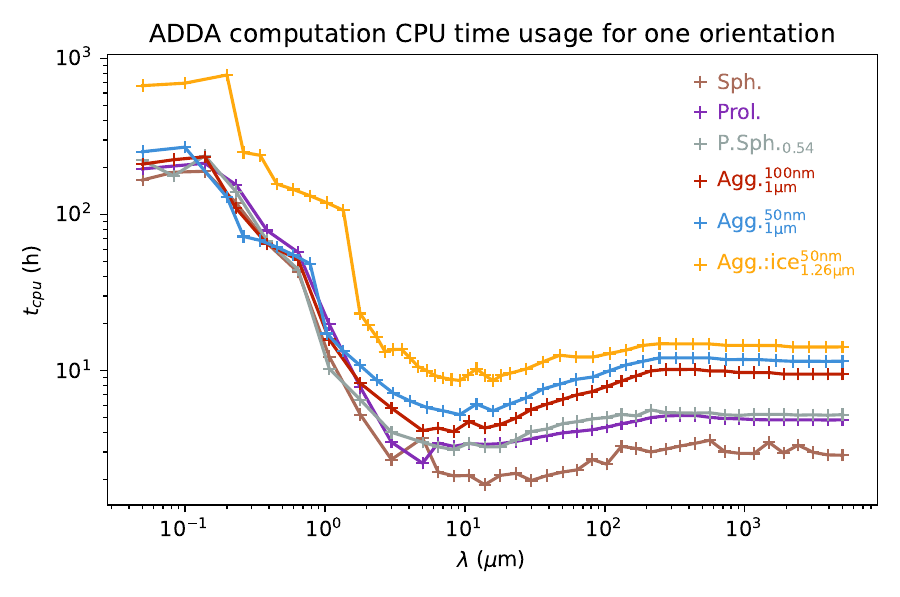}
  \caption{CPU time of the ADDA computations at each wavelength. Times are given for the computation of 1 orientation and total CPU time would need to be multiplied by 19 for the spheroid and 24 for aggregates.}
  \label{fig:CPUtime}
\end{figure}

As explained in details in \cref{subsec:dda}, the modelling of grains requires a great amount of dipoles in the \ac{dda} computations. It means that calculations become really heavy both in terms of required CPU memory and in terms of CPU time. For instance, we need \qty{700}{\hour} (CPU) to compute the optical properties of one orientation of the \aggice at \qty{50}{nm}. All CPU time required for one orientation of the grains for each wavelength are plotted \cref{fig:CPUtime}. This is the reason why we chose to use TGCC resources. Calculations on different orientations and different wavelength are embarrassingly parallel and can be run completely independently with a perfect scaling between the number of CPU and the computing time. However parallelisation of single ADDA computation is a little bit more tricky and requires a little calculation to allocate the right amount of resources.

From \citet[Eq.~(4)]{yurkin_discrete-dipole-approximation_2011}, and given the CPU memory on the TGCC machines $M_{\mathrm{pp}} = \qty{1.8}{Go}$, we can derive the we can derive the necessary number of processes $n_\mathrm{p}$ for ADDA computation by solving the second order equation : 
\begin{equation}
    \left(-M_\mathrm{pp} + 384 n_x^2\right)n_\mathrm{p}^2 + \left(288n_x^3+271(a_\eff/d)^3\right)n_\mathrm{p} + 192n_x^2 = 0,
\end{equation}
where $n_x$ represent the size of the computational box in number of dipoles and $d$ is the size of the dipoles. When the number of dipoles gets too high, there is no more real solution to the equation, meaning one must allocate more than 1 CPU to each task to increase the memory per process $M_{\mathrm{pp}}$. In computations for this study, single computations used between 9 and 32 CPUs at a time.

\section{Particle orientation and scattering direction}
\label{sec:Angles}

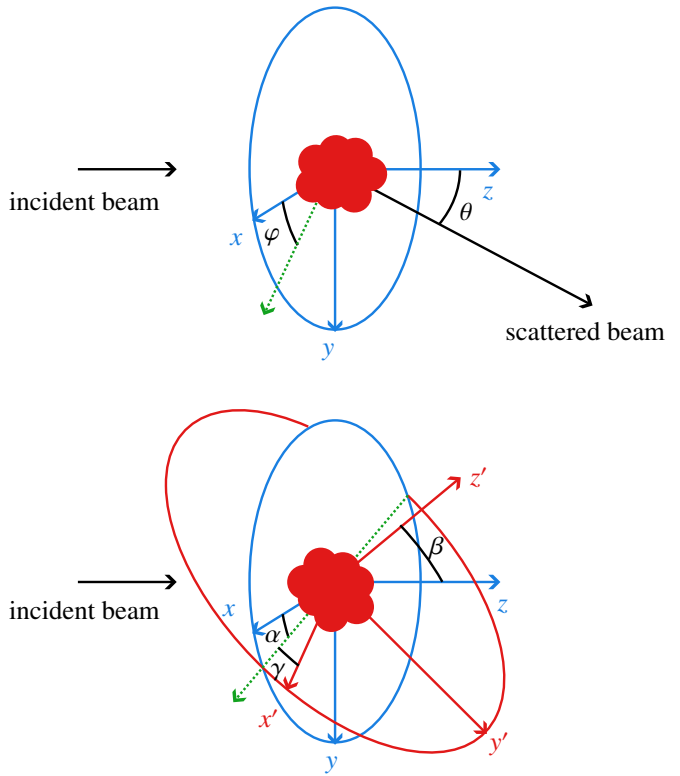
\begin{figure}[h]
  \centering
  \def\svgwidth{\linewidth}

  \begingroup%
      \makeatletter%
      \providecommand\color[2][]{%
        \errmessage{(Inkscape) Color is used for the text in Inkscape, but the package 'color.sty' is not loaded}%
        \renewcommand\color[2][]{}%
      }%
      \providecommand\transparent[1]{%
        \errmessage{(Inkscape) Transparency is used (non-zero) for the text in Inkscape, but the package 'transparent.sty' is not loaded}%
        \renewcommand\transparent[1]{}%
      }%
      \providecommand\rotatebox[2]{#2}%
      \newcommand*\fsize{\dimexpr\f@size pt\relax}%
      \newcommand*\lineheight[1]{\fontsize{\fsize}{#1\fsize}\selectfont}%
      \ifx\svgwidth\undefined%
        \setlength{\unitlength}{557.59434497bp}%
        \ifx\svgscale\undefined%
          \relax%
        \else%
          \setlength{\unitlength}{\unitlength * \real{\svgscale}}%
        \fi%
      \else%
        \setlength{\unitlength}{\svgwidth}%
      \fi%
      \global\let\svgwidth\undefined%
      \global\let\svgscale\undefined%
      \makeatother%
      \begin{picture}(1,1.12855766)%
        \lineheight{1}%
        \setlength\tabcolsep{0pt}%
        \put(0,0){\includegraphics[width=\unitlength,page=1]{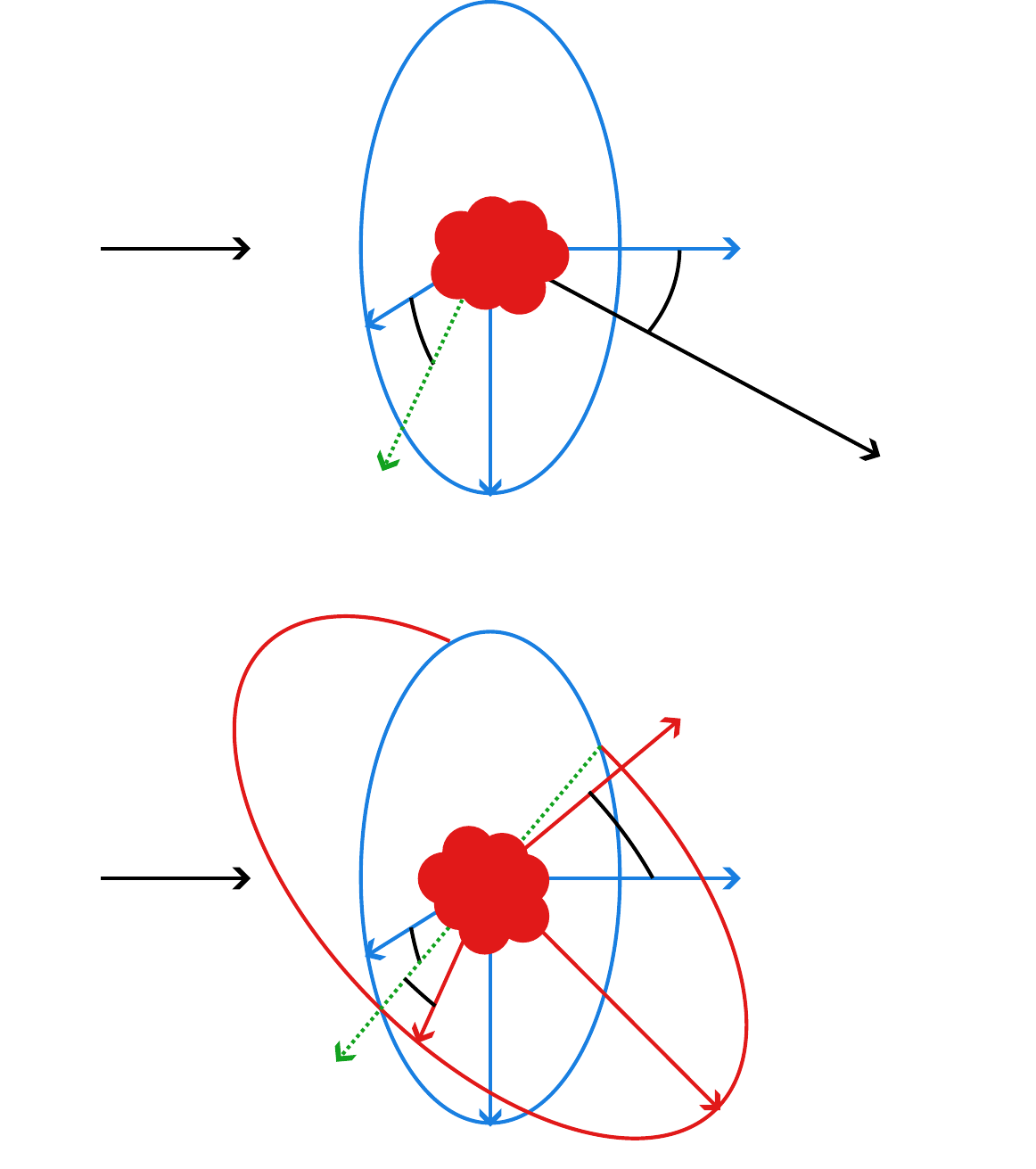}}%
        \put(0.37030301,0.78480838){\color[rgb]{0,0,0}\makebox(0,0)[lt]{\lineheight{1.25}\smash{\begin{tabular}[t]{l}$\varphi$\end{tabular}}}}%
        \put(0.68938184,0.84611762){\color[rgb]{0.09803922,0.49803922,0.88235294}\makebox(0,0)[lt]{\lineheight{1.25}\smash{\begin{tabular}[t]{l}$z$\end{tabular}}}}%
        \put(0.45527469,0.61444112){\color[rgb]{0.09803922,0.49803922,0.88235294}\makebox(0,0)[lt]{\lineheight{1.25}\smash{\begin{tabular}[t]{l}$y$\end{tabular}}}}%
        \put(0.6125485,0.32246609){\color[rgb]{0,0,0}\makebox(0,0)[lt]{\lineheight{1.25}\smash{\begin{tabular}[t]{l}$\beta$\end{tabular}}}}%
        \put(0.37109281,0.19250755){\color[rgb]{0,0,0}\makebox(0,0)[lt]{\lineheight{1.25}\smash{\begin{tabular}[t]{l}$\alpha$\end{tabular}}}}%
        \put(0.65569577,0.81423955){\color[rgb]{0,0,0}\makebox(0,0)[lt]{\lineheight{1.25}\smash{\begin{tabular}[t]{l}$\theta$\end{tabular}}}}%
        \put(0.38020199,0.1430252){\color[rgb]{0,0,0}\makebox(0,0)[lt]{\lineheight{1.25}\smash{\begin{tabular}[t]{l}$\gamma$\end{tabular}}}}%
        \put(-0.00306493,0.22611537){\color[rgb]{0,0,0}\makebox(0,0)[lt]{\lineheight{1.25}\smash{\begin{tabular}[t]{l}incident beam\end{tabular}}}}%
        \put(-0.00306493,0.82870452){\color[rgb]{0,0,0}\makebox(0,0)[lt]{\lineheight{1.25}\smash{\begin{tabular}[t]{l}incident beam\end{tabular}}}}%
        \put(0.72277306,0.63731268){\color[rgb]{0,0,0}\makebox(0,0)[lt]{\lineheight{1.25}\smash{\begin{tabular}[t]{l}scattered beam\end{tabular}}}}%
        \put(0.31883902,0.77730693){\color[rgb]{0.09803922,0.49803922,0.88235294}\makebox(0,0)[lt]{\lineheight{1.25}\smash{\begin{tabular}[t]{l}$x$\end{tabular}}}}%
        \put(0.71090257,0.23814891){\color[rgb]{0.09803922,0.49803922,0.88235294}\makebox(0,0)[lt]{\lineheight{1.25}\smash{\begin{tabular}[t]{l}$z$\end{tabular}}}}%
        \put(0.45999404,0.00721574){\color[rgb]{0.09803922,0.49803922,0.88235294}\makebox(0,0)[lt]{\lineheight{1.25}\smash{\begin{tabular}[t]{l}$y$\end{tabular}}}}%
        \put(0.30774529,0.23114408){\color[rgb]{0.09803922,0.49803922,0.88235294}\makebox(0,0)[lt]{\lineheight{1.25}\smash{\begin{tabular}[t]{l}$x$\end{tabular}}}}%
        \put(0.67077337,0.4231767){\color[rgb]{0.88235294,0.09803922,0.09803922}\makebox(0,0)[lt]{\lineheight{1.25}\smash{\begin{tabular}[t]{l}$z'$\end{tabular}}}}%
        \put(0.69983417,0.03242421){\color[rgb]{0.88235294,0.09803922,0.09803922}\makebox(0,0)[lt]{\lineheight{1.25}\smash{\begin{tabular}[t]{l}$y'$\end{tabular}}}}%
        \put(0.36097767,0.0664491){\color[rgb]{0.88235294,0.09803922,0.09803922}\makebox(0,0)[lt]{\lineheight{1.25}\smash{\begin{tabular}[t]{l}$x'$\end{tabular}}}}%
      \end{picture}%
    \endgroup%@
  
  \caption{Representation of the scattering angles $(\theta,\varphi)$ and the Euler angles $(\alpha,\beta,\gamma)$. In green is represented the line of nodes ($Ox$ after the first rotation). For comprehension purposes, the incident beam has been set to come from the left of the figure, tilting the figures \qty{90}{\degree} to the right from the usual spherical coordinates representation ($Oz$ axis pointing upwards).} 
  \label{fig:angles}
\end{figure}

When computing the optical properties of a dust grain, two sets of angles are of importance. The first angles are referred to as the "orientation of the particle", and actually determine the position of the incident beam of light relatively to the surface of the grain. These angles are classical "Euler angles" $(\alpha,\beta,\gamma)$ representing the orientation of a 3D object in space : $\alpha$ around $Oz$, then $\beta$ around the line of nodes and $\gamma$ around $Oz'$. The second set of angles determines the outward direction of the light, or scattered beam, and has to be set in reference of the incident beam. Two angles are needed to identify the scattered beam : $(\theta,\varphi)$, $\phi$ being the azimuthal angle and $\theta$ the polar angle. All angles are represented \cref{fig:angles}.

\section{Comparison of ice addition effect versus same material addition}
\label{sec:Iceaddidion}

\begin{figure*}[h]
\centering
\begin{subfigure}{.33\textwidth}
  \centering
  \includegraphics[width=\linewidth,clip]{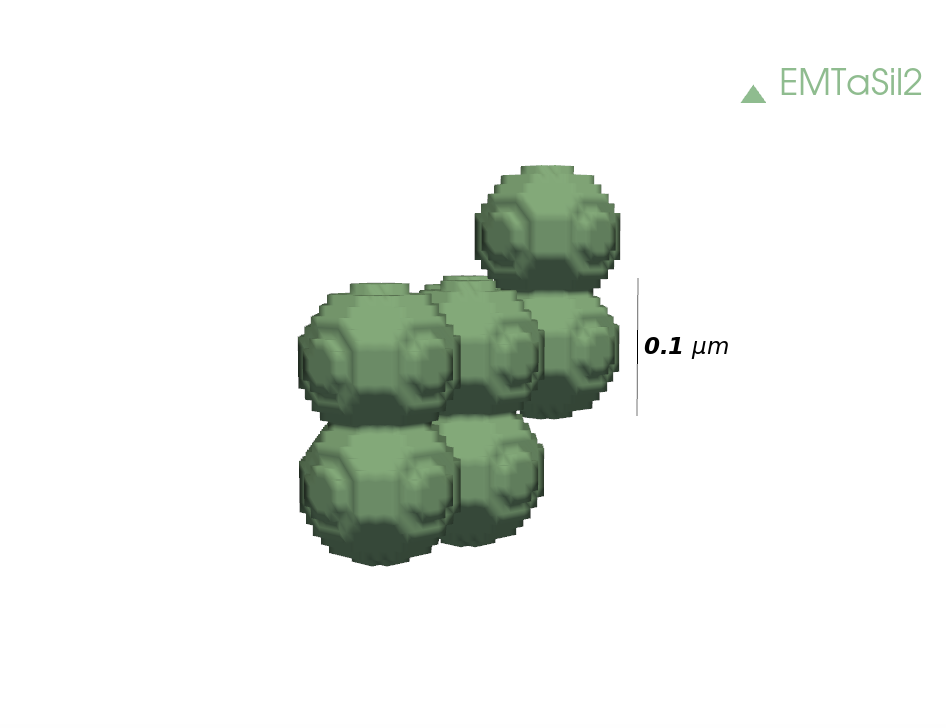}
  \caption{Agg.$^{50 \mathrm{nm}}_{{100}\mathrm{nm}}$}
  \label{fig:sfigA1}
\end{subfigure}%
\begin{subfigure}{.33\textwidth}
  \centering
  \includegraphics[width=\linewidth,clip]{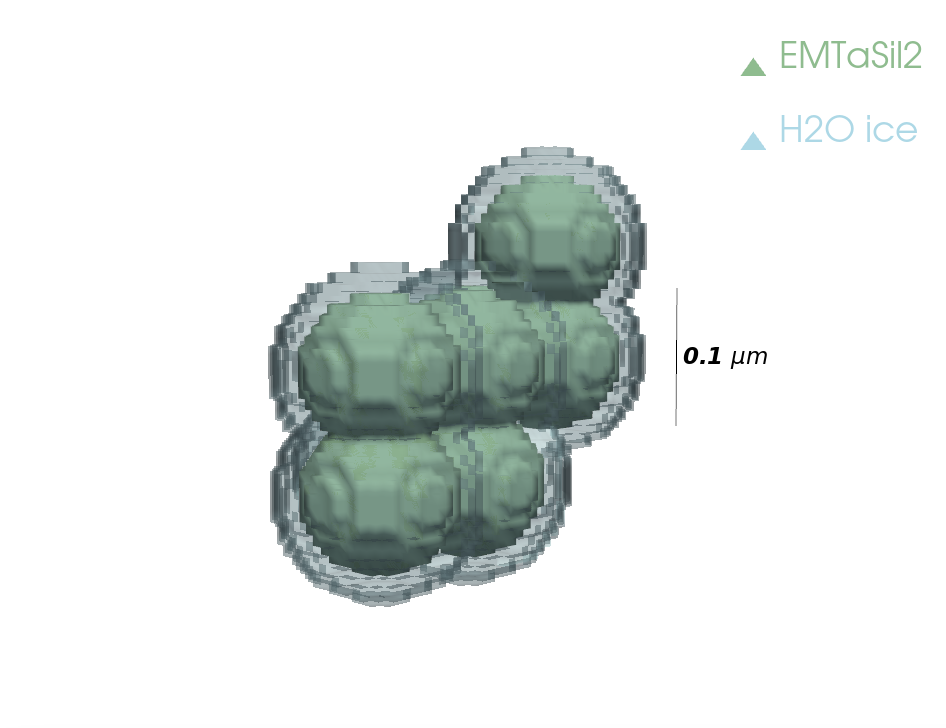}
  \caption{Agg.:ice$^{50 \mathrm{nm}}_{{100}\mathrm{nm}}$}
  \label{fig:sfigA2}
\end{subfigure}%
\begin{subfigure}{.33\textwidth}
  \centering
  \includegraphics[width=\linewidth,clip]{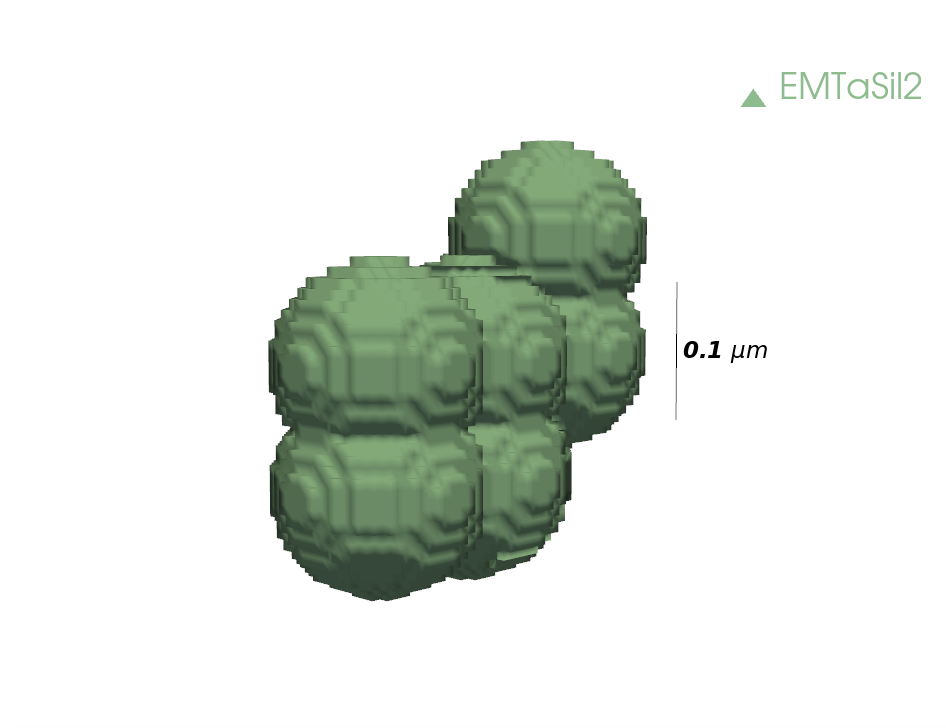}
  \caption{Agg.:a-Sil$^{50 \mathrm{nm}}_{{100}\mathrm{nm}}$}
  \label{fig:sfigA3}
\end{subfigure}
\caption{3D representations of the small aggregates dust grains studied, with dipole size 6.67 nm. All grains are represented with the same scale.}
\label{fig:A3Drep}
\end{figure*}

\begin{figure}[h]
  \centering
  \adjincludegraphics[width=\linewidth,trim={{.52\width} 0 0 {.5\height}},clip]{figures/Qsmall.pdf}
  \caption{Emissivity index for the 3 types of silicate grains. The emissivity index is computed as the slope of the extinction efficiency, and is zoomed on the sub-mm and \acs{mm}.} 
  \label{fig:iceaddbeta}
\end{figure}

\begin{figure}[h]
  \centering
  \includegraphics[width=\linewidth,clip]{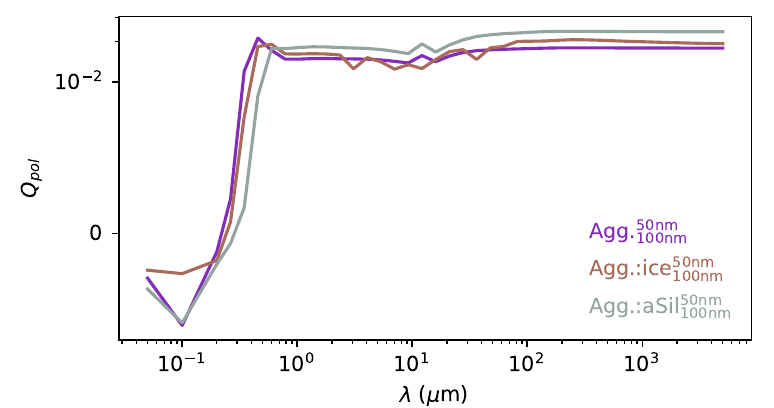}
  \caption{Polarisation efficiency for the 3 types of grains. The y-axis scale is sym-log, meaning that values between $-10^{-2}$ and $10^{-2}$ are plotted on a linear scale to allow the 0 crossing, and the remaining of the axis is in log scale.} 
  \label{fig:iceaddQpol}
\end{figure}

In this section we give a little more detail on the effects created by the addition of an ice mantle on the dust grains. As noted in the paper, this water ice addition generates an increase in the grain total effective radius. To discriminate between effects due to water ice in itself, and effects due to the modification of the grain's geometry when adding this mantle, we conducted further study on small ($a_\eff = \qty{100}{nm}$) aggregates. 

Three grains were considered : a "standard" grain Agg.$^{50 \mathrm{nm}}_{{100}\mathrm{nm}}$, constituted of 8 \qty{50}{nm} monomers, an ice-coated grain Agg.:ice$^{50 \mathrm{nm}}_{{100}\mathrm{nm}}$ created identically to Agg.:ice$^{50 \mathrm{nm}}_{{1}\mathrm{\mu m}}$, and lastly Agg.:a-Sil$^{50 \mathrm{nm}}_{{1}\mathrm{\mu m}}$, which is essentially the same grain, but instead of the ice mantle, we add a mantle of same EMT material as the bulk grain. The three grains are represented \cref{fig:A3Drep}.

The results on emissivity index are plotted \cref{fig:iceaddbeta} and show, as seen in \cref{subsec:mm-ice}, that adding ice mantle to the silicate grains flattens the emissivity. More precisely, we identify here a pure effect of the ice material and no geometric effect : no significant change in the emissivity index is seen in the Agg.:a-Sil$^{50 \mathrm{nm}}_{{1}\mathrm{\mu m}}$ grain compared to Agg.$^{50 \mathrm{nm}}_{{100}\mathrm{nm}}$. However, we cannot say such for polarisation effects, which are likely due to geometrical effects, as polarisation efficiency differs largely from Agg.$^{50 \mathrm{nm}}_{{1}\mathrm{\mu m}}$ to Agg.:a-Sil$^{50 \mathrm{nm}}_{{1}\mathrm{\mu m}}$.

\end{appendix}

\end{document}